\documentclass[nofootinbib,prd,aps,superscriptaddress,preprintnumbers,twocolumn,showpacs]{revtex4-1}
\usepackage[dvipdfmx]{graphicx}
\usepackage{amsmath,amssymb,amsthm,bm,bigdelim,multirow}
\usepackage{color}
\usepackage{booktabs}
\usepackage{dsfont}
\usepackage{braket,cases,latexsym}
\usepackage{here}
\newcommand{\im}{\mathrm{i}}

\newcommand{\gTr}{\mathrm{gTr}}
\newcommand{\e}{\mathrm{e}}

\newcommand{\lo}{\mathrm{O}}

\begin{document}

\title{
Matrix product decomposition for two- and three-flavor Wilson fermions: Benchmark results in the lattice Gross-Neveu model at finite density
}

\author{Shinichiro Akiyama}
	\email[]{akiyama@phys.s.u-tokyo.ac.jp}
	\affiliation{Graduate School of Science, University of Tokyo, Tokyo, 113-003, Japan}

\begin{abstract}
We formulate the path integral of two- and three-flavor Wilson fermion in two dimensions as a multilayer Grassmann tensor network by the matrix product decomposition. 
Thanks to this new description, the memory cost scaling is reduced from $\lo(\e^{N_{f}})$ for the conventional construction to $\lo(N_{f})$.
Based on this representation, we develop a coarse-graining algorithm where spatially or temporally adjacent Grassmann tensors are converted into a canonical form along a virtual direction before we carry out the spacetime coarse-graining. 
Benchmarking with the lattice Gross-Neveu model at finite density, we see that the Silver Blaze phenomenon in the pressure and number density is captured with relatively small bond dimensions.
\end{abstract}

\maketitle

\section{Introduction}
\label{sec:intro}

The tensor network method has been widely applied in many fields including condensed matter physics, information science, and particle physics~\cite{Orus:2018dya,Banuls:2019rao,Banuls:2019bmf,Meurice:2020pxc,Okunishi:2021but,Kadoh:2022loj}. 
In the context of the quantum many-body problem, the tensor network method can be classified namely by two streams.
One is the method based on the Hamiltonian formalism, where we express a quantum state as a certain form of tensor contraction and try to optimize each tensor by some variational methods, such as the density matrix renormalization group (DMRG)~\cite{White:1992zz,White:1993zza}.
The other is based on the Lagrangian formalism, where we represent the path integral as a tensor contraction and approximately perform the contraction employing a variant of the real-space renormalization group method, such as the Levin-Nave tensor renormalization group (TRG)~\cite{Levin:2006jai}. 
One of the most remarkable aspects of the latter approach is that several TRG algorithms enable us to deal with an infinitely large lattice even in higher dimensions if systems are translationally invariant on a lattice~\cite{PhysRevB.86.045139,Adachi:2019paf,Kadoh:2019kqk}.
Since the usual bottleneck part of the higher-dimensional TRG calculation is the tensor contraction, parallel computation techniques play a significant role in the practical applications~\cite{Akiyama:2019chk,Yamashita:2021yxs}.
So far, the TRG approach has been applied to several four-dimensional lattice theories~\cite{Akiyama:2020ntf,Akiyama:2020soe,Milde:2021vln,Akiyama:2022eip}. 

When we try to apply the TRG approach to systems with larger internal degrees of freedom, however, one encounters several difficulties. 
The size of each tensor in its tensor network representation increases and a larger memory space is required in the numerical calculation.
The bond dimension, which characterizes the maximal size of the tensor under the tensor network method, may have to be enlarged to reduce an approximation to the original representation as much as possible.
To resolve these issues, one of the promising approaches is to find an efficient expression for such systems in the language of tensor networks.
Typical examples are non-Abelian lattice gauge theories, where naive derivations of their tensor network representation, particularly in four dimensions, requires extremely large memory costs.
Recently, several attempts have been made for two-dimensional theories~\cite{Bazavov:2019qih,Fukuma:2021cni,Hirasawa:2021qvh,Luo:2022eje} and the three-dimensional $SU(2)$ pure gauge theory~\cite{Kuwahara:2022ubg}. 
These works have successfully demonstrated the applicability of the TRG approach for lattice theories with non-Abelian degrees of freedom.
On the other hand, one also encounters a difficulty originating from larger internal spaces when flavor degrees of freedom are introduced to lattice fermions.
This can be another challenging issue because we have to deal with a local tensor, whose size scales exponentially with respect to $N_{f}$, the total number of flavors. 
So far, single-flavor lattice fermions have been usually employed in the previous TRG studies~\cite{Shimizu:2014uva,Shimizu:2014fsa,Takeda:2014vwa,Shimizu:2017onf,PhysRevD.101.094509,Akiyama:2020soe,Bloch:2022vqz}.
\footnote{Recently, the loop-TNR algorithm~\cite{PhysRevLett.118.110504} has been applied to the two-flavor lattice Gross-Neveu model on a small lattice in Ref.~\cite{Asaduzzaman:2022pnw}.}
Therefore, how to deal with finite-$N_{f}$ systems with the TRG approach needs to be addressed, particularly for its future application toward the $(2+1)$-flavor quantum chromodynamics (QCD) on a lattice, for instance.
To this aim, we are going to apply a tensor network decomposition for the Grassmann tensor network representation~\cite{Akiyama:2020sfo} from the beginning and formulate its coarse-graining procedure.
The main idea is based on the so-called matrix product decomposition, which has been very familiar as the matrix product state (MPS)~\cite{PhysRevLett.75.3537,Dukelsky_1998} within the Hamiltonian formalism.
\footnote{Generalization of the MPS toward higher-dimensional systems has been actively attempted~\cite{PhysRevE.64.016705,Verstraete:2004cf,PhysRevLett.124.037201,Dai:2022oks}, including an application to the four-dimensional lattice quantum electrodynamics at finite density~\cite{Magnifico:2020bqt}.}
We numerically test our algorithm, employing the Gross-Neveu model~\cite{Gross:1974jv} at finite density with the two- and three-flavor Wilson fermions.
\footnote{The model has been employed to benchmark the world line and fermion bag approaches~\cite{Ayyar:2017xmi} and quantum simulation~\cite{Asaduzzaman:2022bpi}, recently. See also the references therein.}
Using the matrix product decomposition, the model with $N_{f}=2$ and $N_{f}=3$ can be described as two- and three-layer Grassmann tensor network whose bond dimension is equal to four.
Thanks to this new description, the memory cost scaling is reduced from $\lo(\e^{N_{f}})$ for the conventional construction to $\lo(N_{f})$.
Based on this description, we develop a coarse-graining procedure, by which we calculate pressure and number density to verify whether the Silver Blaze phenomenon~\cite{Cohen:2003kd} can be captured or not.

This paper is organized as follows. 
In Sec.~\ref{sec:method}, we introduce the matrix product decomposition of the Grassmann tensor for the lattice Gross-Neveu model with the Wilson fermion. 
Based on this representation, we propose a coarse-graining procedure, taking advantage of matrix product decomposition, particularly of canonical form.
In Sec.~\ref{sec:results}, we first check the efficiency of the matrix product decomposition for the initial Grassmann tensor network. 
We secondly perform free field computation, where the free-energy density as a function of volume is calculated by our method and a naive TRG application, and show our method successfully reproduces the exact solution. 
Thirdly, we present numerical results of pressure and number density as functions of chemical potential with the finite coupling constant. 
As the validation, we also employ the bond-weighted TRG (BTRG)~\cite{PhysRevB.105.L060402} to evaluate the original tensor network representation.
Section~\ref{sec:summary} is devoted to summary and outlook.

\section{Algorithm}
\label{sec:method}

To explain our algorithm, we consider the Gross-Neveu-Wilson (GNW) model with $N_{f}=3$ at finite density on a square lattice $\Lambda_{2}=\{n=(n_{1},n_{2})|n_{\nu}\in\mathds{Z},~\nu=1,2\}$. 
The lattice action is
\begin{widetext}
\begin{align}
\label{eq:action}
	S
	&=
	-\frac{1}{2}\sum_{n\in\Lambda_{2}}\sum_{\nu=1,2}\sum_{f=1}^{N_{f}}
	\left[
		\e^{\mu\delta_{\nu,2}}\bar{\psi}^{(f)}_{n}(r\mathds{1}-\gamma_{\nu})\psi^{(f)}_{n+\hat{\nu}}
		+\e^{-\mu\delta_{\nu,2}}\bar{\psi}^{(f)}_{n+\hat{\nu}}(r\mathds{1}+\gamma_{\nu})\psi^{(f)}_{n}
	\right]
	\nonumber\\
	&+\sum_{n}\sum_{f=1}(M+2r)\bar{\psi}^{(f)}_{n}\psi^{(f)}_{n}
	-\frac{g^{2}_{\sigma}}{2N_{f}}\sum_{n}
	\left(\sum_{f}\bar{\psi}^{(f)}_{n}\psi^{(f)}_{n}\right)^{2}
	-\frac{g^{2}_{\pi}}{2N_{f}}\sum_{n}
	\left(\sum_{f}\bar{\psi}^{(f)}_{n}\im\gamma_{5}\psi^{(f)}_{n}\right)^{2},
\end{align}
\end{widetext}
where $\psi^{(f)}_{n}$ and $\bar{\psi}^{(f)}_{n}$ describe the Wilson fermions, labeled by the flavor index $f$, which are two-component Grassmann-valued fields.
$g^{2}_{\sigma}$ and $g^{2}_{\pi}$ are coupling constants. 
$M$, $\mu$, and $r$ represent mass, chemical potential, and the Wilson parameter, respectively.
In this study, we always assume that these parameters do not depend on $f$.
In addition, we always set $r=1$ and employ the Pauli matrices to represent the two-dimensional $\gamma$-matrix via $\gamma_{1}=\sigma_{x}$, $\gamma_{2}=\sigma_{y}$, and $\gamma_{5}=-\im\gamma_{1}\gamma_{2}=\sigma_{z}$. 
We regard $\nu=1~(2)$ as the spatial (temporal) direction, assuming the (anti-)periodic boundary condition. 

\subsection{Matrix product decomposition at initial stage}
\label{subsec:initial}

According to Ref.~\cite{Akiyama:2020sfo}, the tensor network representation of the path integral $Z$ generated by the lattice action~\eqref{eq:action} is expressed in the form of
\begin{align}
\label{eq:path_integral}
	Z=\gTr\left[\prod_{\mathrm{lattice~site}}\mathcal{T}\right],
\end{align}
where $\mathcal{T}$ is a Grassmann tensor, which is a multilinear combination of Grassmann numbers. 
\footnote{TRG approach for fermions was firstly introduced in Refs.~\cite{Gu:2010yh,Gu:2013gba}}
The size of the Grassmann tensor is determined by the lattice geometry and the hopping structure in Eq.\eqref{eq:action}. 
We have four adjacent sites for each site on the square lattice $\Lambda_{2}$ and forward and backward hopping terms in each direction, for each flavor. 
Therefore, $\mathcal{T}$ is written in the following way,
\begin{widetext}
\begin{align}
\label{eq:g_tensor}
	\mathcal{T}
	&=
	%\nonumber\\
	\left(\prod_{f=1}^{N_{f}}
	\sum_{x_{1}^{(f)}, x_{2}^{(f)}, t_{1}^{(f)}, t_{2}^{(f)}, x_{1}^{(f)'}, x_{2}^{(f)'}, t_{1}^{(f)'}, t_{2}^{(f)'}}
	\right)
	T_{
	x_{1}^{(1)}x_{2}^{(1)}\cdots x_{1}^{(N_{f})}x_{2}^{(N_{f})}
	t_{1}^{(1)}t_{2}^{(1)}\cdots t_{1}^{(N_{f})}t_{2}^{(N_{f})}
	x_{1}^{(1)'}x_{2}^{(1)'}\cdots x_{1}^{(N_{f})'}x_{2}^{(N_{f})'}
	t_{1}^{(1)'}t_{2}^{(1)'}\cdots t_{1}^{(N_{f})'}t_{2}^{(N_{f})'}
	}
	\nonumber\\
	&\times
	\eta^{x_{1}^{(1)}x_{2}^{(1)}\cdots x_{1}^{(N_{f})}x_{2}^{(N_{f})}}
	\xi^{t_{1}^{(1)}t_{2}^{(1)}\cdots t_{1}^{(N_{f})}t_{2}^{(N_{f})}}
	\bar{\eta}^{t_{2}^{(N_{f})'}t_{1}^{(N_{f})'}\cdots t_{2}^{(1)'}t_{1}^{(1)'}}
	\bar{\xi}^{x_{2}^{(N_{f})'}x_{1}^{(N_{f})'}\cdots x_{2}^{(1)'}x_{1}^{(1)'}}
	,
\end{align}
\end{widetext}
where $T$ is referred to as the coefficient tensor of $\mathcal{T}$ whose indices take $0$ or $1$ and we set
\begin{align}
\label{eq:eta}
	&\eta^{x_{1}^{(1)}x_{2}^{(1)}\cdots x_{1}^{(N_{f})}x_{2}^{(N_{f})}}
	\nonumber\\
	&=
	(\eta_{1}^{(1)})^{x_{1}^{(1)}}(\eta_{2}^{(1)})^{x_{2}^{(1)}}
	\cdots
	(\eta_{1}^{(N_{f})})^{x_{1}^{(N_{f})}}(\eta_{2}^{(N_{f})})^{x_{2}^{(N_{f})}}
	,
\end{align}
\begin{align}
\label{eq:xi}
	&\xi^{t_{1}^{(1)}t_{2}^{(1)}\cdots t_{1}^{(N_{f})}t_{2}^{(N_{f})}}
	\nonumber\\
	&=
	(\xi_{1}^{(1)})^{t_{1}^{(1)}}(\xi_{2}^{(1)})^{t_{2}^{(1)}}
	\cdots
	(\xi_{1}^{(N_{f})})^{t_{1}^{(N_{f})}}(\xi_{2}^{(N_{f})})^{t_{2}^{(N_{f})}}
	,
\end{align}
\begin{align}
\label{eq:bar_eta}
	&\bar{\eta}^{t_{2}^{(N_{f})'}t_{1}^{(N_{f})'}\cdots t_{2}^{(1)'}t_{1}^{(1)'}}
	\nonumber\\
	&=
	(\bar{\eta}_{2}^{(N_{f})})^{x_{2}^{(N_{f})'}}(\bar{\eta}_{1}^{(N_{f})})^{x_{1}^{(N_{f})'}}
	\cdots
	(\bar{\eta}_{2}^{(1)})^{x_{2}^{(1)'}}(\bar{\eta}_{1}^{(1)})^{x_{1}^{(1)'}}
	,
\end{align}
\begin{align}
\label{eq:bar_xi}
	&\bar{\xi}^{x_{2}^{(N_{f})'}x_{1}^{(N_{f})'}\cdots x_{2}^{(1)'}x_{1}^{(1)'}}
	\nonumber\\
	&=
	(\bar{\xi}_{2}^{(N_{f})})^{t_{2}^{(N_{f})'}}(\bar{\xi}_{1}^{(N_{f})})^{t_{1}^{(N_{f})'}}
	\cdots
	(\bar{\xi}_{2}^{(1)})^{t_{2}^{(1)'}}(\bar{\xi}_{1}^{(1)})^{t_{1}^{(1)'}}
	.
\end{align}
$\eta_{i}^{(f)}$, $\xi_{i}^{(f)}$, $\bar{\eta}_{i}^{(f)}$, $\bar{\xi}_{i}^{(f)}$ ($i=1,2$) are single-component auxiliary Grassmann numbers. 
Note that the ordering of Grassmann numbers in Eq.~\eqref{eq:bar_eta} (Eq.~\eqref{eq:bar_xi}) is different from Eq.~\eqref{eq:eta} (Eq.~\eqref{eq:xi}). 
These orderings help us to integrate auxiliary Grassmann numbers in Eq.~\eqref{eq:path_integral} under the coarse-graining procedure~\cite{Akiyama:2020sfo}. 
The explicit form of coefficient tensor in Eq.~\eqref{eq:g_tensor} can be straightforwardly obtained by the procedure explained in Ref.~\cite{Akiyama:2020sfo}.

Since the coefficient tensor $T$ consists of $4^{4N_{f}}$ elements, the size of the Grassmann tensor in Eq.~\eqref{eq:g_tensor} increases exponentially with respect to $N_{f}$. 
We now decompose the coefficient tensor into the lower-rank tensors, each of which is related to each flavor's degree of freedom. 
Firstly, we rearrange indices of $T$ and auxiliary Grassmann numbers in Eq.~\eqref{eq:g_tensor} as
\begin{widetext}
\begin{align}
\label{eq:reordered_T}
\mathcal{T}
	&=
	%\nonumber\\
	\left(\prod_{f=1}^{N_{f}}
	\sum_{x_{1}^{(f)}, x_{2}^{(f)}, t_{1}^{(f)}, t_{2}^{(f)}, x_{1}^{(f)'}, x_{2}^{(f)'}, t_{1}^{(f)'}, t_{2}^{(f)'}}
	\right)
	T_{
	x_{1}^{(1)}x_{2}^{(1)}t_{1}^{(1)}t_{2}^{(1)}x_{1}^{(1)'}x_{2}^{(1)'}t_{1}^{(1)'}t_{2}^{(1)'}
	\cdots 
	x_{1}^{(N_{f})}x_{2}^{(N_{f})}t_{1}^{(N_{f})}t_{2}^{(N_{f})}x_{1}^{(N_{f})'}x_{2}^{(N_{f})'}t_{1}^{(N_{f})'}t_{2}^{(N_{f})'}
	}
	\nonumber\\
	&\times
	\Phi^{x_{1}^{(1)}x_{2}^{(1)}t_{1}^{(1)}t_{2}^{(1)}x_{2}^{(1)'}x_{1}^{(1)'}t_{2}^{(1)'}t_{1}^{(1)'}}
	\cdots
	\Phi^{x_{1}^{(N_{f})}x_{2}^{(N_{f})}t_{1}^{(N_{f})}t_{2}^{(N_{f})}x_{2}^{(N_{f})'}x_{1}^{(N_{f})'}t_{2}^{(N_{f})'}t_{1}^{(N_{f})'}}
	,
\end{align}
\end{widetext}
where we have assumed that an extra sign factor coming from the rearrangement of Grassmann numbers is included in $T$. $\Phi$ describes 
\begin{align}
	&\Phi^{x_{1}^{(f)}x_{2}^{(f)}t_{1}^{(f)}t_{2}^{(f)}x_{2}^{(f)'}x_{1}^{(f)'}t_{2}^{(f)'}t_{1}^{(f)'}}
	\nonumber\\
	&=
	(\eta_{1}^{(f)})^{x_{1}^{(f)}}(\eta_{2}^{(f)})^{x_{2}^{(f)}}
	(\xi_{1}^{(f)})^{t_{1}^{(f)}}(\xi_{2}^{(f)})^{t_{2}^{(f)}}
	\nonumber\\
	&\times
	(\bar{\eta}_{2}^{(f)})^{x_{2}^{(f)'}}(\bar{\eta}_{1}^{(f)})^{x_{1}^{(f)'}}
	(\bar{\xi}_{2}^{(f)})^{t_{2}^{(f)'}}(\bar{\xi}_{1}^{(f)})^{t_{1}^{(f)'}}
	.
\end{align}
Defining an eight-bit string $I_{f}$ as
\begin{align}
	I_{f}
	=
	x_{1}^{(f)}x_{2}^{(f)}t_{1}^{(f)}t_{2}^{(f)}x_{1}^{(f)'}x_{2}^{(f)'}t_{1}^{(f)'}t_{2}^{(f)'}
	,
\end{align}
we consider the matrix product decomposition for $T$ in Eq.~\eqref{eq:reordered_T}, which is easily obtained by repeating the singular value decomposition (SVD) of $T$ twice,
\begin{align}
\label{eq:mpd_nf3}
	&T_{I_{1}I_{2}I_{3}}
	=
	\sum_{\alpha_{1}=1}^{\chi}
	U_{I_{1}\alpha_{1}}\left(\sigma_{\alpha_{1}}V^{*}_{I_{2}I_{3}\alpha_{1}}\right)
	\nonumber\\
	&=
	\sum_{\alpha_{1}=1}^{\chi}
	\sum_{\alpha_{2}=1}^{\chi}
	U_{I_{1}\alpha_{1}}
	%\left(
	U_{\alpha_{1}I_{2}\alpha_{2}}\sigma_{\alpha_{2}}V^{*}_{I_{3}\alpha_{2}}
	%\right)
	.
\end{align}
Note that $U$'s and $V$'s are isometries and $\sigma$ denotes the singular value. 
Throughout this paper, we distinguish tensors by their subscripts.
$\chi$ is referred to as the flavor bond dimension.
The right-hand side of Eq.~\eqref{eq:mpd_nf3} is known as the canonical form, where all tensors in the matrix product decomposition satisfy some orthogonality conditions. In case of Eq.~\eqref{eq:mpd_nf3}, we have orthogonality conditions such that
\begin{align*}
	\sum_{I_{1}}
	U_{I_{1}\alpha_{1}}U^{*}_{I_{1}\alpha'_{1}}
	=
	\delta_{\alpha_{1}\alpha'_{1}}
	,
\end{align*}
\begin{align*}
	\sum_{I_{2},\alpha_{1}}
	U_{\alpha_{1}I_{2}\alpha_{2}}U^{*}_{\alpha_{1}I_{2}\alpha'_{2}}
	=
	\delta_{\alpha_{2}\alpha'_{2}}
	,
\end{align*}
\begin{align*}
	\sum_{I_{3}}
	V^{*}_{I_{3}\alpha_{2}}V_{I_{3}\alpha'_{2}}
	=
	\delta_{\alpha_{2}\alpha'_{2}}
	.
\end{align*}
In the tensor network calculation based on the Hamiltonian formalism, one uses the MPS with its canonical form to represent a state vector.
One of the advantages of the canonical form is that one can calculate physical quantities on a larger system just via some local manipulations.
For a detailed explanation of the canonical form, see Ref.~\cite{Schollw_ck_2011}, for instance.
In the following, we will see that the canonical form enables us to calculate some reduced density matrices easily and to develop a coarse-graining procedure of the Grassmann tensor network via some local operations with respect to each flavor.
The matrix product decomposition of Eq.~\eqref{eq:mpd_nf3} can be rewritten as 
\begin{align}
\label{eq:vfc_nf3}
	T_{I_{1}I_{2}I_{3}}
	=
	\sum_{\alpha_{1}=1}^{\chi}
	\sum_{\alpha_{2}=1}^{\chi}
	\Gamma^{I_{1}}_{\alpha_{1}}
	\sigma_{\alpha_{1}}
	\Gamma^{I_{2}}_{\alpha_{1}\alpha_{2}}
	\sigma_{\alpha_{2}}
	\Gamma^{I_{3}}_{\alpha_{2}}
	,
\end{align}
where $\Gamma^{I_{1}}_{\alpha_{1}}=U_{I_{1}\alpha_{1}}$, $\Gamma^{I_{2}}_{\alpha_{1}\alpha_{2}}=\sigma_{\alpha_{1}}^{-1}U_{\alpha_{1}I_{2}\alpha_{2}}$, and $\Gamma^{I_{3}}_{\alpha_{2}}=V^{*}_{I_{3}\alpha_{2}}$.
The right-hand side of Eq.~\eqref{eq:vfc_nf3} is a different type of canonical form which was first proposed in Ref.~\cite{PhysRevLett.91.147902}, in the context of the classical simulation of the quantum computations based on the Hamiltonian formalism.
In this canonical form, singular values exist explicitly between all $\Gamma$ tensors.
For Eqs.~\eqref{eq:mpd_nf3} and \eqref{eq:vfc_nf3} to hold, a naive choice of $\chi$ is $\chi=2^{8}$. 
If we have the vanishing singular values, however, we have a chance to compress original coefficient tensors by expressing the right-hand sides of Eqs.~\eqref{eq:mpd_nf3} and \eqref{eq:vfc_nf3}.
We will see later that $\chi=4$ is sufficiently large for these equations to hold.

\begin{figure*}[htbp]
	%\centering
	\includegraphics[width=1.0\hsize]{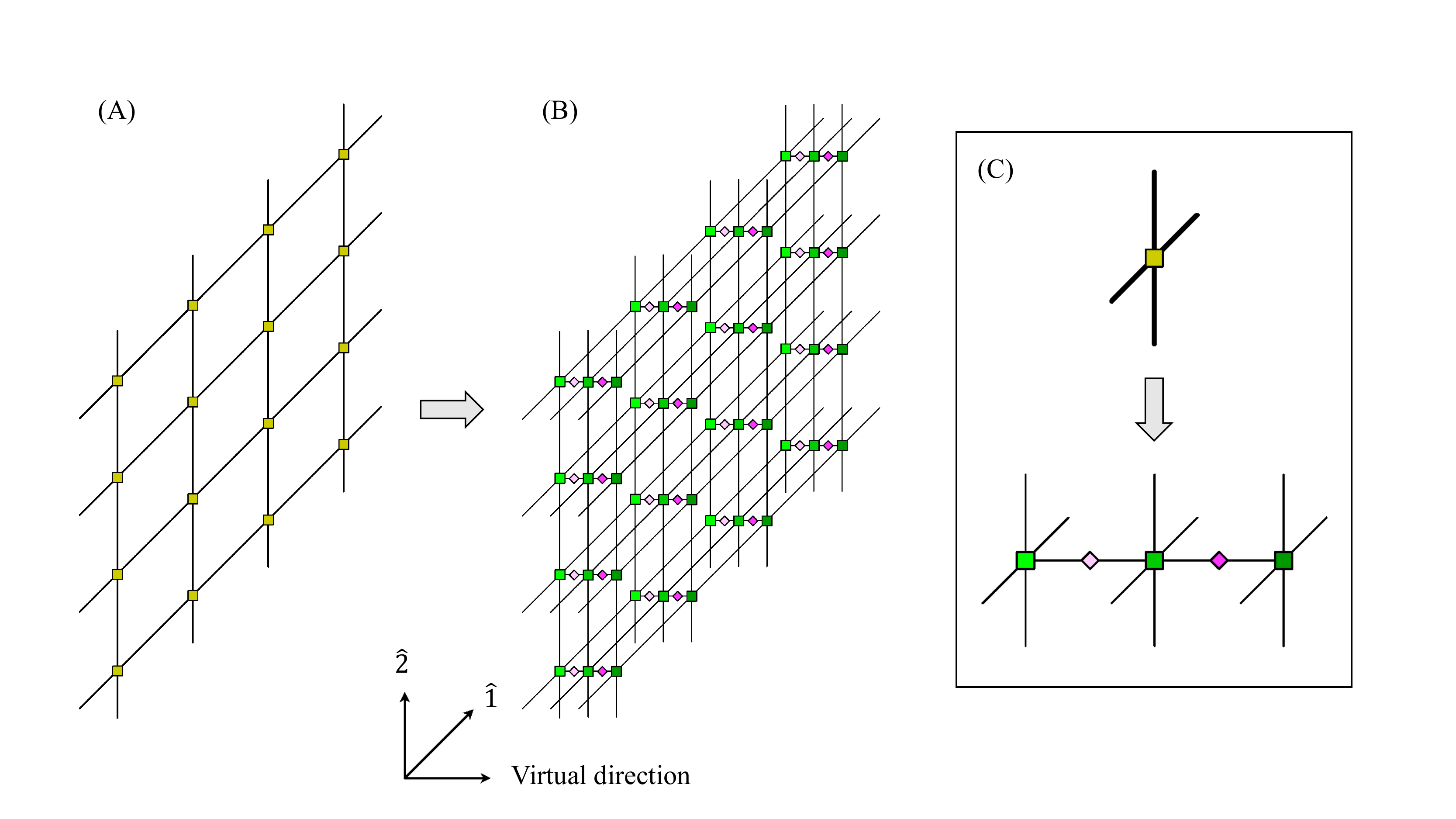}
	\caption{
	Schematic illustration of Eq.~\eqref{eq:path_integral} with $N_{f}=3$. 
	(A) Original description of Eq.~\eqref{eq:path_integral}. 
	Yellow square symbols show the Grassmann tensor defined by Eq.~\eqref{eq:g_tensor}. 
	(B) Equation~\eqref{eq:path_integral} generated by Eq.~\eqref{eq:vfc_g_nf3}. 
	Green square symbols show $\mathcal{A}_{\alpha_{1}}$, $\mathcal{B}_{\alpha_{1}\alpha_{2}}$ and $\mathcal{C}_{\alpha_{2}}$ in Eq.~\eqref{eq:vfc_g_nf3}.
	Diamond symbols correspond to $\sigma_{\alpha_{1}}$ and $\sigma_{\alpha_{2}}$ in Eq.~\eqref{eq:vfc_g_nf3}.
	Contractions parallel to the original two-dimensional plane are given by the Grassmann contractions. 
	On the other hand, contractions along the virtual direction are given by normal tensor contractions.
	(C) Diagrammatic representation of Eq.~\eqref{eq:vfc_g_nf3}.
	}
  	\label{fig:mpd}
\end{figure*}

Obviously, the matrix product decomposition of the coefficient tensor can be interpreted as that of the corresponding Grassmann tensor. 
Eq.~\eqref{eq:vfc_nf3} provides us with 
\begin{align}
\label{eq:vfc_g_nf3}
	\mathcal{T}
	=
	\sum_{\alpha_{1}=1}^{\chi}
	\sum_{\alpha_{2}=1}^{\chi}
	\mathcal{A}_{\alpha_{1}}
	\sigma_{\alpha_{1}}
	\mathcal{B}_{\alpha_{1}\alpha_{2}}
	\sigma_{\alpha_{2}}
	\mathcal{C}_{\alpha_{2}}
	,
\end{align}
where $\mathcal{A}_{\alpha_{1}}$, $\mathcal{B}_{\alpha_{1}\alpha_{2}}$ and $\mathcal{C}_{\alpha_{2}}$ are the Grassmann tensors, defined by contracting $I_{f}$ ($f=1,2,3$) with corresponding Grassmann numbers in Eq.~\eqref{eq:reordered_T}, labeled by some extra indices, $\alpha_{1}$, $\alpha_{2}$. 
Figure~\ref{fig:mpd} schematically shows Eq.~\eqref{eq:path_integral} with $N_{f}=3$ and its matrix product decomposition of Eq.~\eqref{eq:vfc_g_nf3}. 
We can understand that Eq.~\eqref{eq:vfc_g_nf3} has introduced a virtual direction for the original two-dimensional Grassmann tensor network.
Notice that the contractions parallel to the original two-dimensional plane are the Grassmann contractions, but those along the virtual direction are normal tensor contractions.

\subsection{Coarse-graining procedure}
\label{subsec:cg}

\begin{figure*}[htbp]
	\centering
	\includegraphics[width=1.0\hsize]{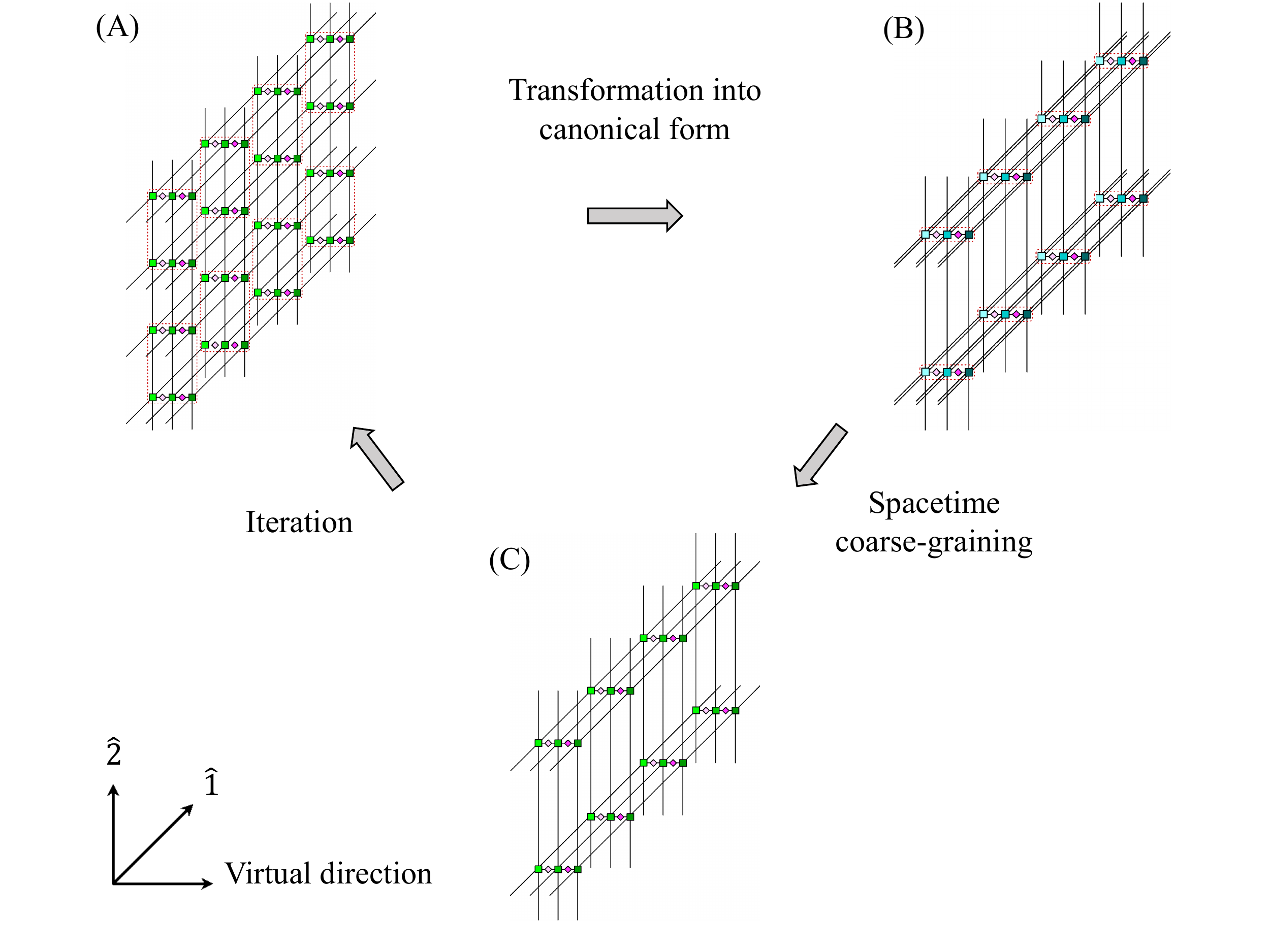}
	\caption{
	Schematic illustration of our coarse-graining procedure. 
	(A) Two adjacent $\mathcal{T}^{(q)}$'s along the temporal (spatial) direction are transformed into a canonical form along the virtual direction with the finite flavor bond dimension $\chi$.
	(B) Spacetime coarse-graining decimates the degrees of freedom along the spatial (temporal) direction with the finite spacetime bond dimension $D$.
	(C) Since the resulting network has the same geometry as it did, we can repeat the procedure.
	}
  	\label{fig:schematic}
\end{figure*}

Our method is basically similar to the algorithm of higher-order TRG (HOTRG)~\cite{PhysRevB.86.045139}.
At each coarse-graining step, however, we convert two adjacent coefficient tensors, along the spatial or temporal direction, into a canonical form along the virtual direction before we carry out the spacetime coarse-graining, doing the same with the HOTRG.
A schematic picture is shown in Fig.~\ref{fig:schematic}.
In addition to the flavor bond dimension $\chi$, we also introduce the spacetime bond dimension $D$ to decimate the degrees of freedom under the coarse-graining.
Let $q$ be an integer to specify the iteration number of coarse-graining. 
Then the algorithm sequentially defines a coarse-graining transformation from two adjacent $\mathcal{T}^{(q)}$'s to $\mathcal{T}^{(q+1)}$, where
\begin{align}
\label{eq:q}
	\mathcal{T}^{(q)}
	=
	\sum_{\alpha_{1}=1}^{\chi}
	\sum_{\alpha_{2}=1}^{\chi}
	\mathcal{A}_{\alpha_{1}}^{(q)}
	\sigma_{\alpha_{1}}^{(q)}
	\mathcal{B}_{\alpha_{1}\alpha_{2}}^{(q)}
	\sigma_{\alpha_{2}}^{(q)}
	\mathcal{C}_{\alpha_{2}}^{(q)}
	.
\end{align}
The coarse-grained tensor $\mathcal{T}^{(q)}$ is defined on a lattice with $2^{q}$ sites.
$\mathcal{T}^{(0)}$ is given by Eq.~\eqref{eq:vfc_g_nf3}.
In other words, repeating this coarse-graining procedure $q$ times, $2^{q}$ numbers of $\mathcal{T}^{(0)}$ are approximately contracted.
We shall start with the coarse-graining along the temporal direction, so we have $2^{k}$ sites in the spatial direction and $2^{k+1}$ sites in the temporal direction when $q=2k+1~(k=0,1,\cdots)$.

To obtain the canonical form with the fixed flavor bond dimension $\chi$, we derive several squeezers (flavor squeezers) as explained in Appendix~\ref{subsec:flavor_cg}.
The SVD carried out by these flavor squeezers updates $\sigma_{\alpha_{1}}^{(q)}$ and $\sigma_{\alpha_{2}}^{(q)}$.
On the other hand, the spacetime coarse-graining is given by different squeezers (spacetime squeezers) with the spacetime bond dimension $D$ as explained in Appendix~\ref{subsec:sp_cg}.
By these spacetime and flavor squeezers, $\mathcal{A}^{(q)}$, $\mathcal{B}^{(q)}$ and $\mathcal{C}^{(q)}$ are updated.
We emphasize that the canonical form allows us to perform the spacetime coarse-graining for each flavor segment, $\mathcal{A}^{(q)}$, $\mathcal{B}^{(q)}$, $\mathcal{C}^{(q)}$, separately.
Furthermore, this coarse-graining scheme can be straightforwardly extended to arbitrary $N_{f}$, keeping the number of local tensors needed to describe the Grassmann tensor network constant, $N_{f}$, at all coarse-graining steps.

\section{Numerical results} 
\label{sec:results}

We provide several benchmark results in the lattice GNW model at finite density.
In the following, we set $g^{2}=g^{2}_{\sigma}=g^{2}_{\pi}$ for simplicity. 
The lattice volume is denoted by $V$.
Firstly, we check the efficiency of the matrix product decomposition at the initial stage: How large should $\chi$ in Eq.~\eqref{eq:vfc_nf3} be to restore the original coefficient tensor?
Secondly, we compute the path integral with $N_{f}=3$ and $g^{2}=0$. 
This computation allows us to verify our coarse-graining procedure via a comparison of the numerical result and the exact solution.
Then, we employ the current algorithm to calculate the GNW model with $g^{2}\neq0$ in the finite chemical potential regime.

\subsection{Matrix product decomposition at initial stage}

\begin{figure}[htbp]
	\centering
	\begin{tabular}{cc}
		\begin{minipage}[t]{\hsize}
			%\centering
			\includegraphics[width=0.75\hsize]{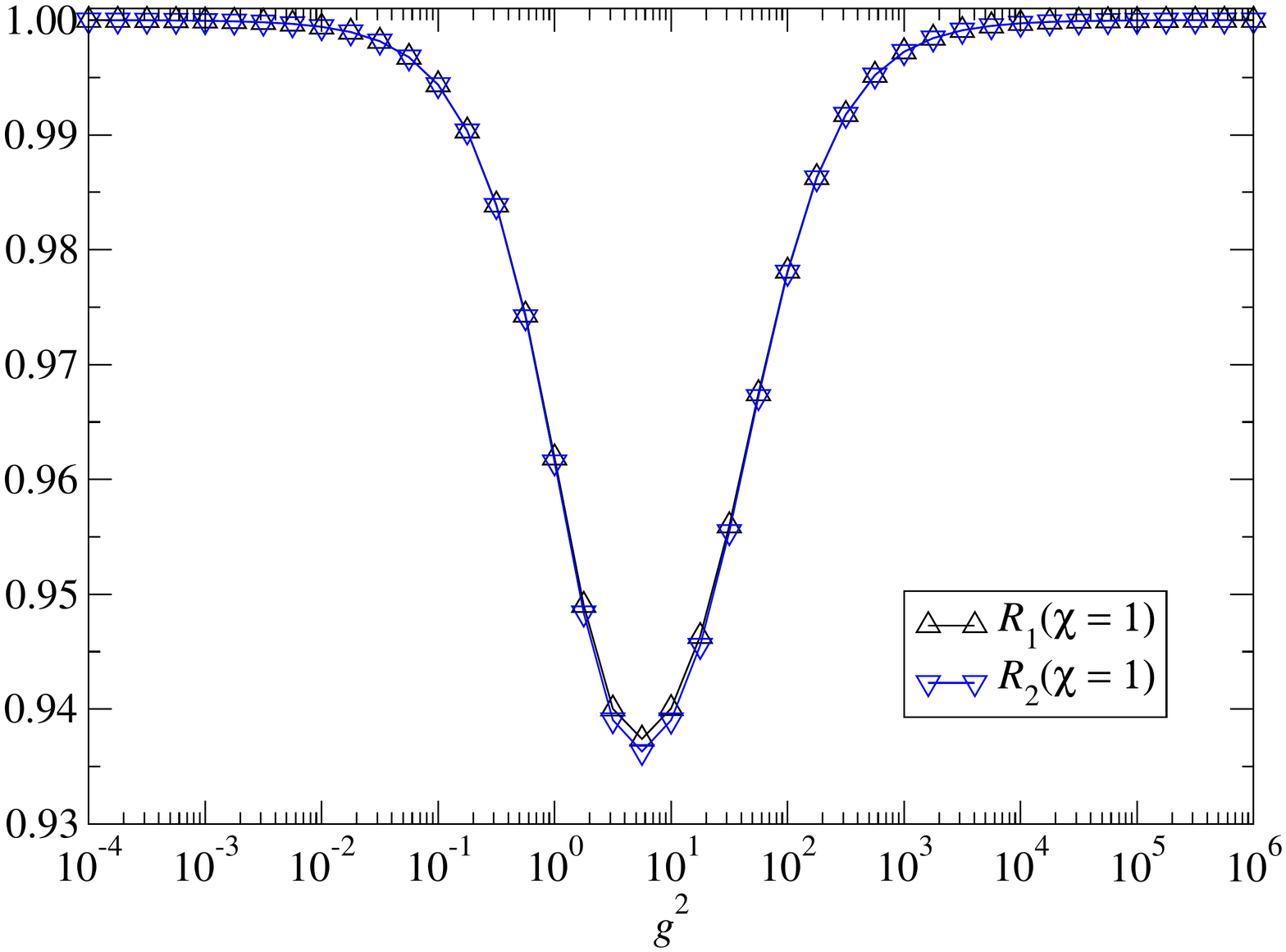}
			%\subcaption{$\chi=1$}
		\end{minipage} \\
		\\
		\begin{minipage}[t]{\hsize}
			%\centering
			\includegraphics[width=0.75\hsize]{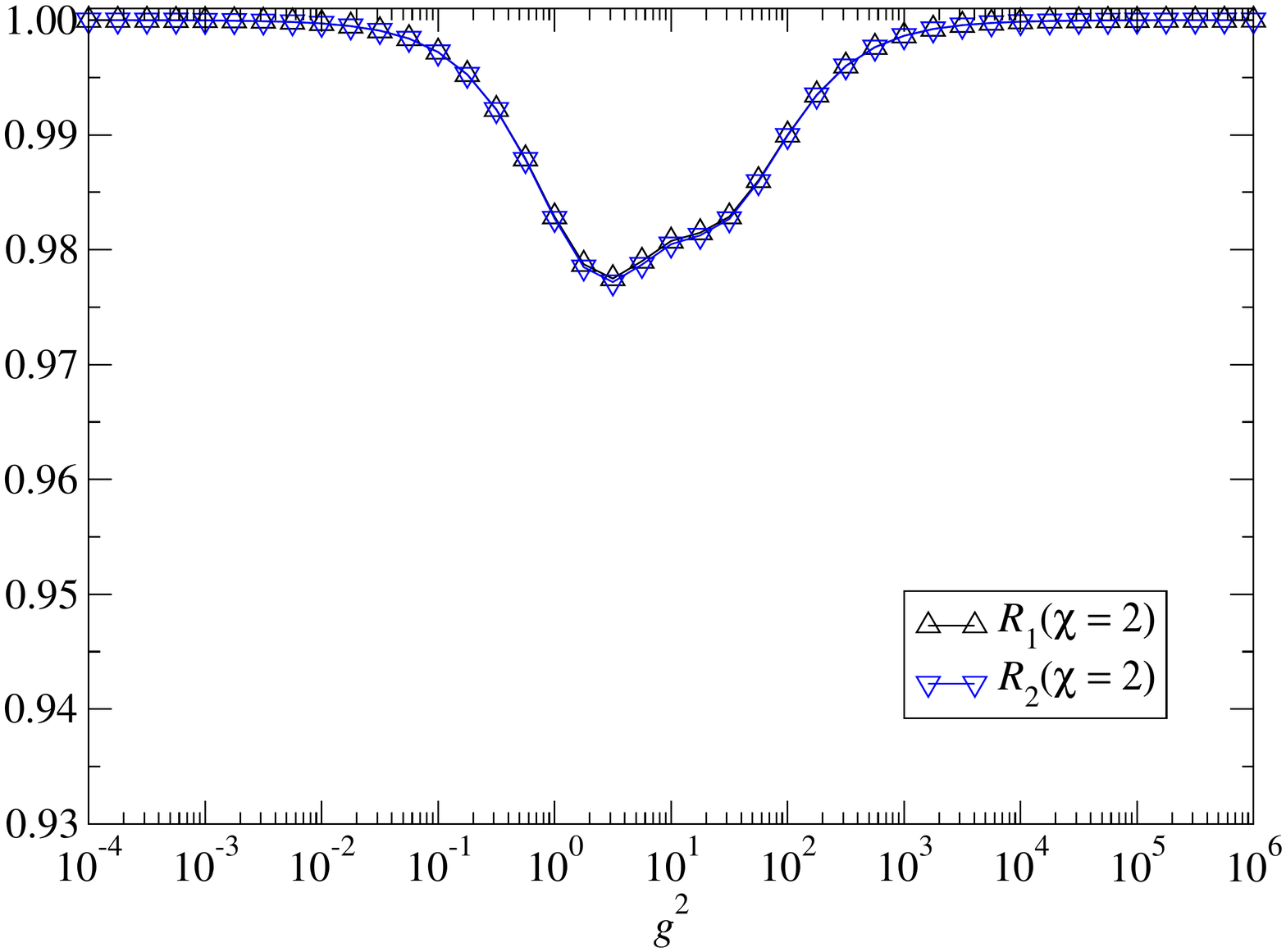}
			%\subcaption{$\chi=2$}
		\end{minipage} \\
		\\
		\begin{minipage}[t]{\hsize}
			%\centering
			\includegraphics[width=0.75\hsize]{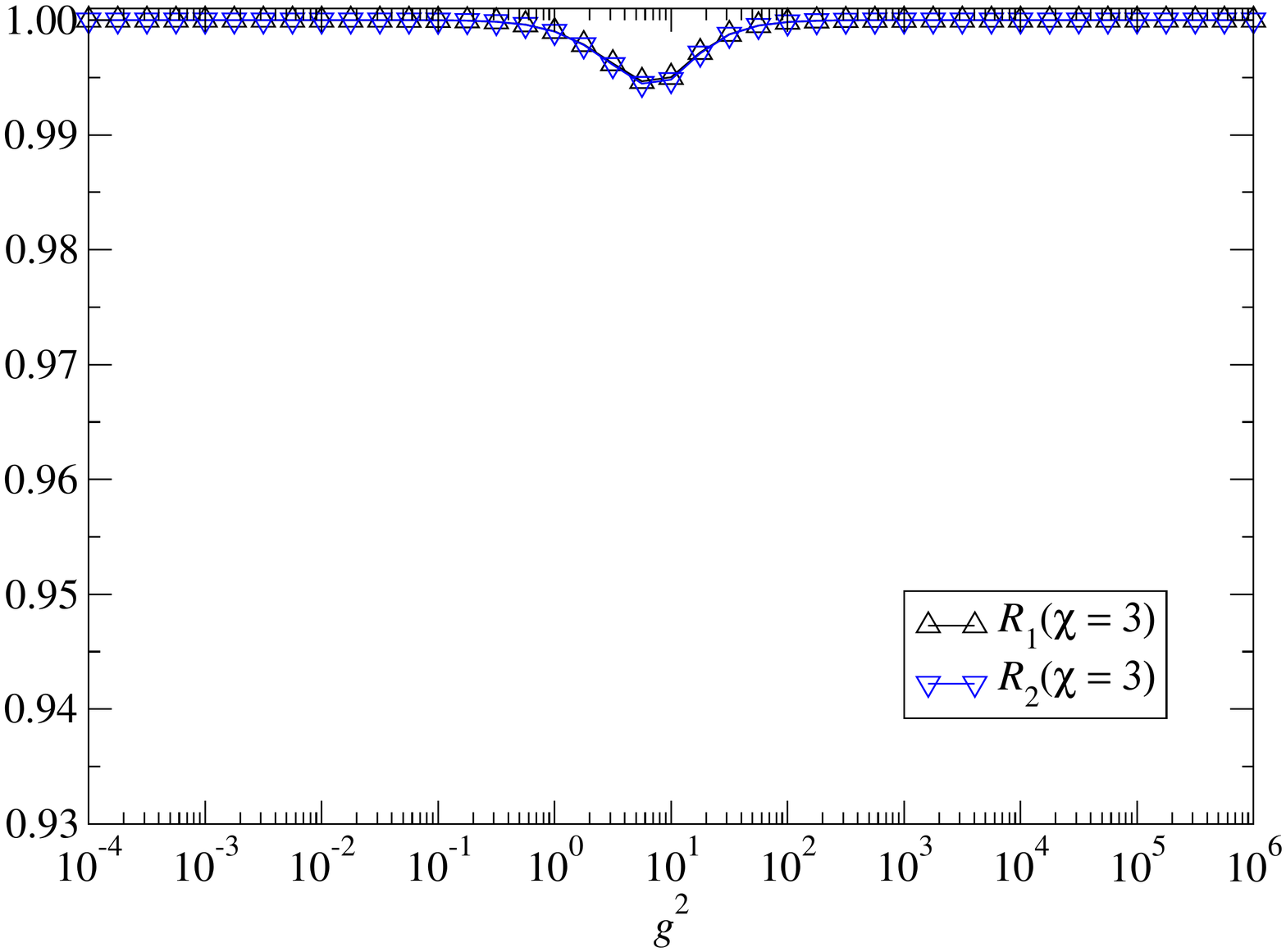}
			%\subcaption{$\chi=3$}
		\end{minipage} \\
		\\
		\begin{minipage}[t]{\hsize}
			%\centering
			\includegraphics[width=0.75\hsize]{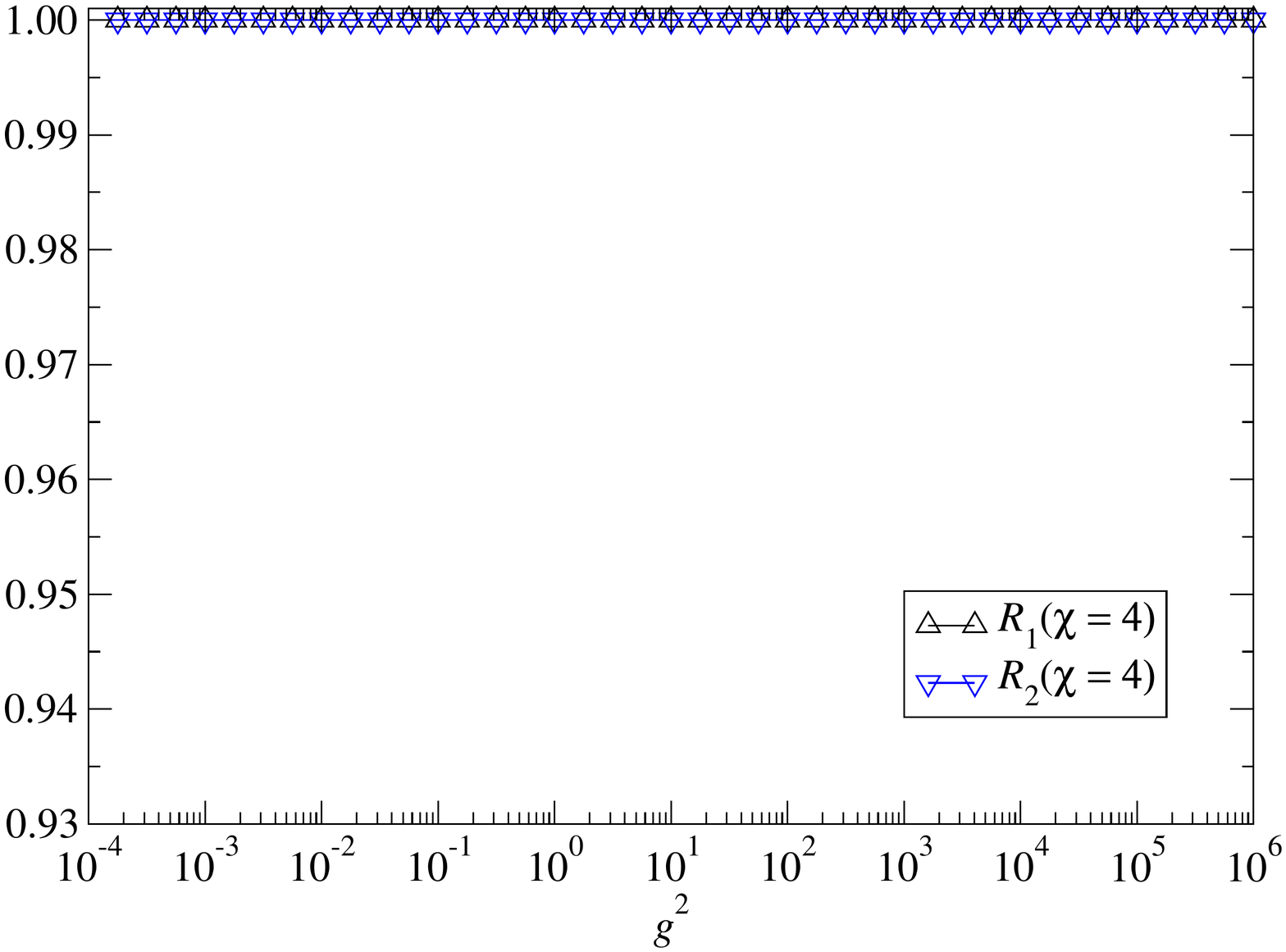}
			%\subcaption{$\chi=4$}
		\end{minipage}
	\end{tabular}
	\caption{$R_{p}(\chi)$ as a function of $g^{2}$ with $\chi=1,2,3,4$. $M$ and $\mu$ are set to be zero.}
	\label{fig:vidal_chi}
\end{figure}

To check the efficiency of the matrix product decomposition at the initial stage, we define
\begin{align}
	R_{p}(\chi)=||T^{(0)}||^{-1}\sqrt{\sum_{\alpha_{p}=1}^{\chi}\sigma^{2}_{\alpha_{p}}}
	,
\end{align}
with $p=1,2$. $||T^{(0)}||$ denotes the Frobenius norm of the initial coefficient tensor and $\sigma_{\alpha_{p}}$'s are the singular values in Eq.~\eqref{eq:vfc_nf3}. 
By definition, $R_{p}(\chi)$ is always less than or equal to $1$. Obviously, $R_{p}(2^{8})=1$ holds. 
Figure~\ref{fig:vidal_chi} shows $R_{p}(\chi)$ as a function of $g^{2}$, varying $\chi$.
We set $M=0$ and $\mu=0$.
We can see that $\chi=4<2^8$ is sufficiently large to make Eq.~\eqref{eq:vfc_nf3} hold.
This means that the two-dimensional GNW model with $N_{f}=3$ can be exactly expressed as the three-layer Grassmann tensor network, whose bond dimension is four, as shown in Fig.~\ref{fig:mpd}~(B).
This can be attributed to the fact that although each $I_{j}~(j=1,2,3)$ in Eq.~\eqref{eq:vfc_nf3} describes $2^{8}$ patterns of combinations of auxiliary Grassmann numbers, most of them have vanishing contribution due to the Grassmann nature. 
Therefore, we expect that this kind of matrix product decomposition is efficient not only for the GNW model but also for other pure fermionic systems.

We can predict that the effect of truncation by flavor squeezers in the coarse-graining procedure should depend on $g^{2}$ because Fig.~\ref{fig:vidal_chi} tells us the singular value spectrum depends on the coupling constant.
Note that $g^{2}=0$ is a special point: when $g^{2}=0$, Eq.~\eqref{eq:vfc_nf3} holds just with $\chi=1$, because the coefficient tensor for $N_{f}$-flavor free fermions is given by the direct product of the $N_{f}$ number of coefficient tensors for the single-flavor free fermion. 

\subsection{Free field computation}

\begin{figure}[htbp]
	\centering
	\includegraphics[width=0.9\hsize]{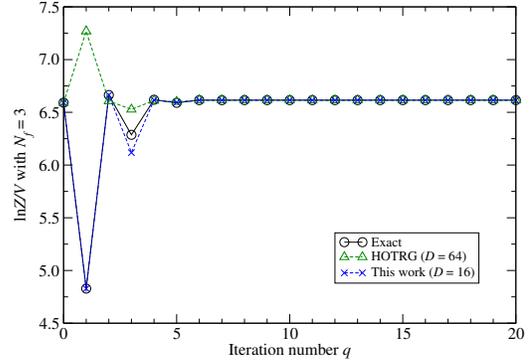}
	\caption{
	Free energy of the three-flavor Wilson fermion with $M=1$ and $\mu=0$, as a function of the iteration number $q$. Circles show the exact solution. Triangles are the result of naive HOTRG with $D=64$. The result obtained by our algorithm is shown as cross symbols.
	When $q$ is odd, it corresponds to the coarse-graining along the temporal direction.
	}
  	\label{fig:nf3_g0m1_trg_history}
\end{figure}

\begin{figure}[htbp]
	\centering
	\includegraphics[width=0.9\hsize]{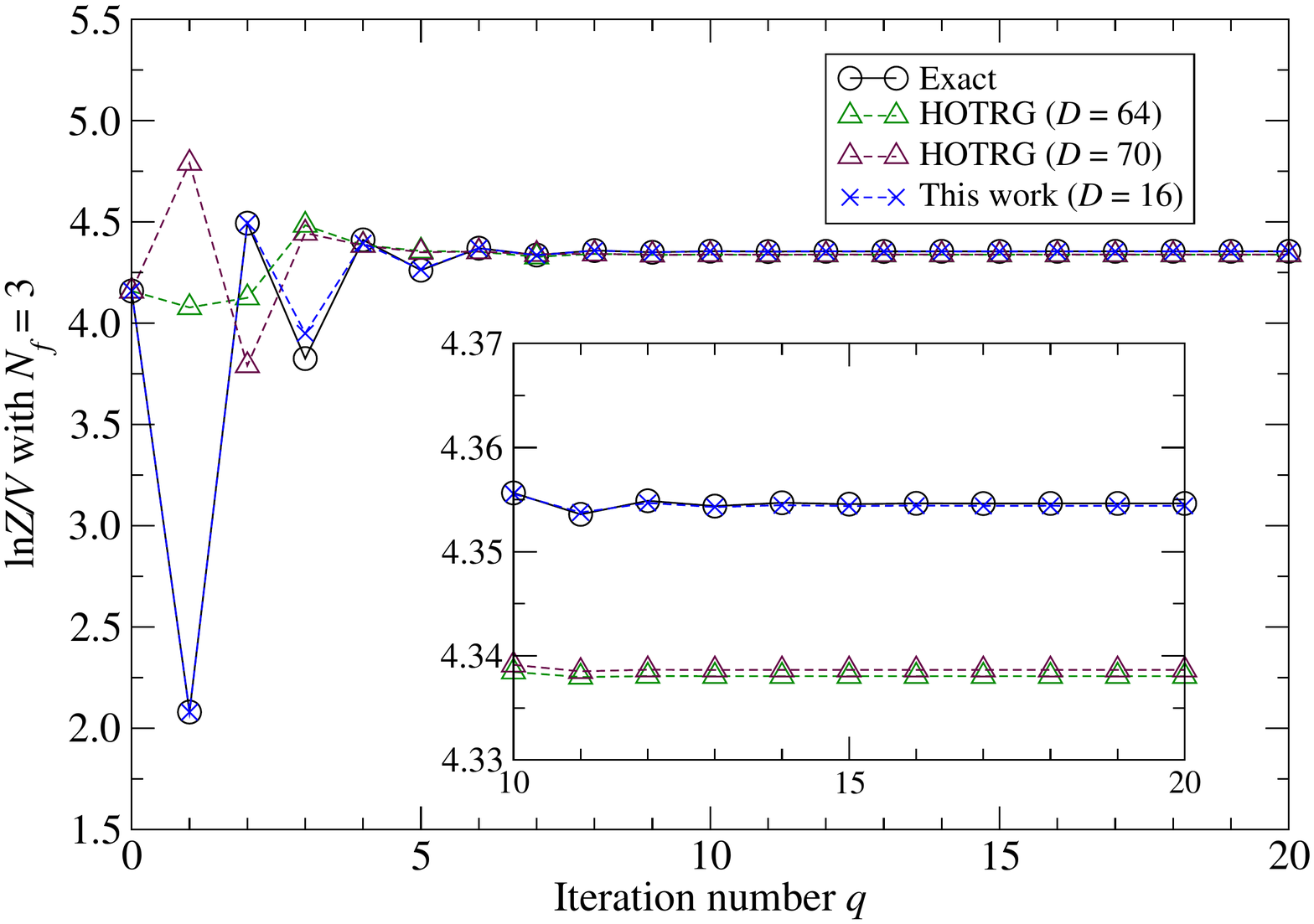}
	\caption{
	Free energy of the three-flavor Wilson fermion with $M=0$ and $\mu=0$, as a function of the iteration number $q$. Circles show the exact solution. Triangles are the results of naive HOTRG with $D=64$ (green) and $D=70$ (brown). The result obtained by our algorithm is shown as cross symbols.
	When $q$ is odd, it corresponds to the coarse-graining along the temporal direction.}
  	\label{fig:nf3_g0_trg_history}
\end{figure}

Next, we calculate three-flavor Wilson fermions with the vanishing interaction. 
As mentioned previously, there is no truncation error coming from the finite flavor bond dimension, because we just need $\chi=1$, so we can focus on the finite-$D$ effect originating from the spacetime squeezers.
Figure~\ref{fig:nf3_g0m1_trg_history} shows the free energy density with $M=1$ and $\mu=0$ as a function of $q$, the iteration number of the coarse-graining introduced in Eq.~\eqref{eq:q}.
It shows that the numerical result by our algorithm with $D=16$ successfully reproduces the exact solution. 
At $V=2^{20}$, the relative error of the free energy is $\lo(10^{-7})$. 
We also have naively employed the HOTRG algorithm with $D=64$, to evaluate the original Grassmann tensor network generated by Eq.~\eqref{eq:g_tensor}.
\footnote{The HOTRG calculations in Figs.~\ref{fig:nf3_g0m1_trg_history} and \ref{fig:nf3_g0_trg_history} are performed not by isometries but by spacetime squeezers. Note that this type of algorithmic extension has already appeared in Refs.~\cite{Iino:2019vxd,Yoshiyama_2020,Adachi:2019paf,https://doi.org/10.48550/arxiv.2303.07733}.}
Although the relative error becomes $\lo(10^{-5})$ at $V=2^{20}$, the small volume results are inconsistent with exact ones.
Note that we have to take $D\ge64$ to avoid an extra approximation for the initial tensor itself in the naive computation.
Since the path integral of the free field theory with multiflavors is given by the product of free field theories with one flavor, the HOTRG can achieve much higher accuracy if one applies it to the single-flavor theory.
However, such a treatment is available only for free field theories.

Similarly, Fig.~\ref{fig:nf3_g0_trg_history} shows the free energy density with $M=0$ and $\mu=0$.
The relative error at $V=2^{20}$ is $\lo(10^{-4})$ by our method.
Again, we have naively employed the HOTRG algorithm with $D=64$ and $D=70$ to evaluate the original Grassmann tensor network.
Their coarse-graining histories suggest that there is little hope to restore the exact solution by the naive HOTRG computations for massless Wilson fermions. 
On the other hand, our method enables us to exactly carry out the tensor contractions up to $V=2^{2}$ regardless of $M$, and to evaluate the path integral in the thermodynamic limit.
Therefore, these numerical results imply that our algorithm could achieve better performance in multiflavor fermion models with relatively small bond dimensions compared with conventional TRG algorithms.

\subsection{Interacting fermions at finite density}

\begin{figure}[htbp]
	\centering
	\includegraphics[width=0.9\hsize]{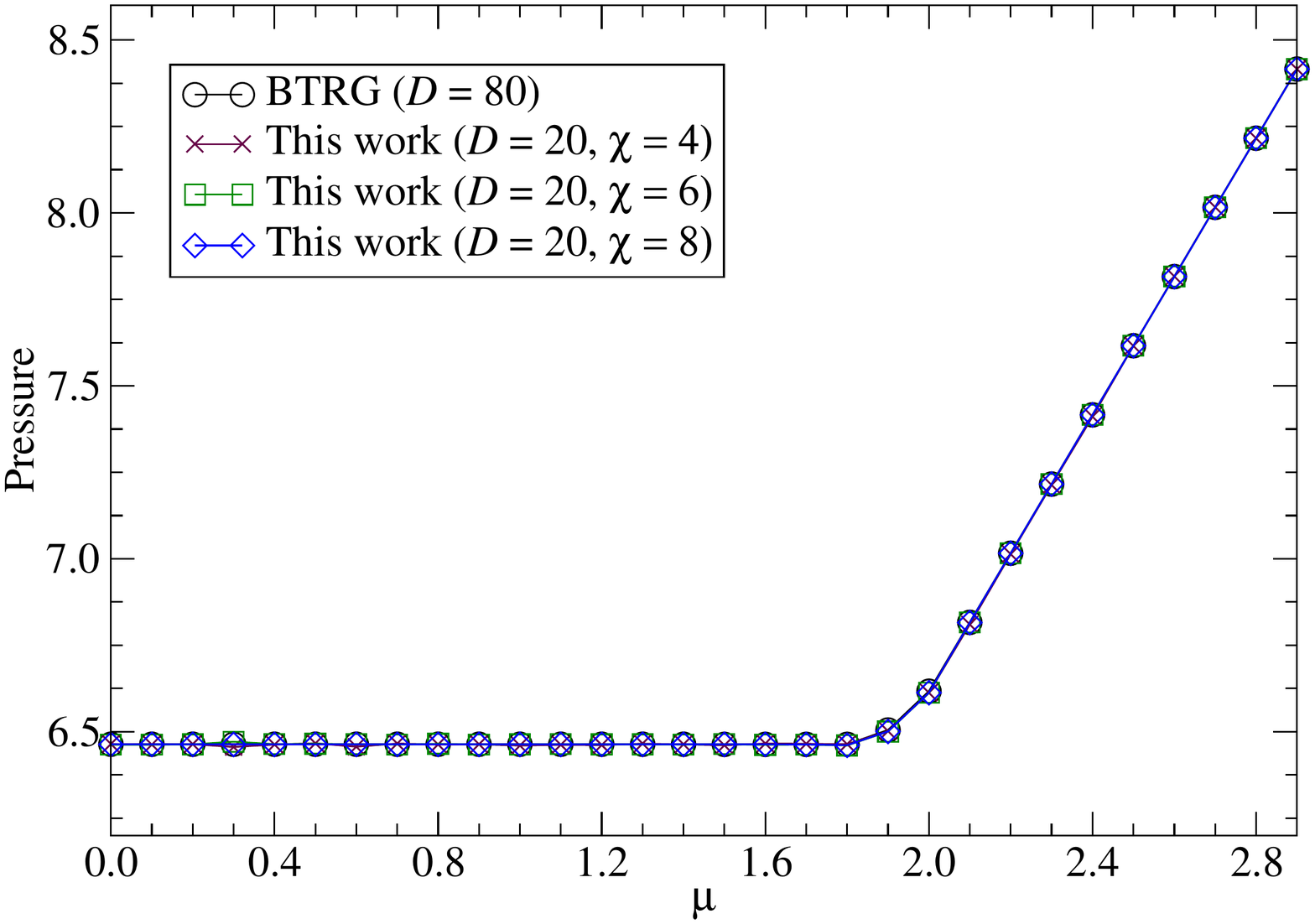}
	\caption{
	Pressure as a function of chemical potential with $N_{f}=2$, $g^2=10$, and $M=1$. 
	Cross, square, and diamond symbols correspond to computations with our algorithm setting $\chi=4,6,8$, respectively.
	The spacetime bond dimension is set as $D=20$ in all cases.
	Circles show the BTRG calculations, setting $D=80$.
	}
	\label{fig:nf2_g10}
\end{figure}
\begin{figure}[htbp]
	\centering
	\includegraphics[width=0.9\hsize]{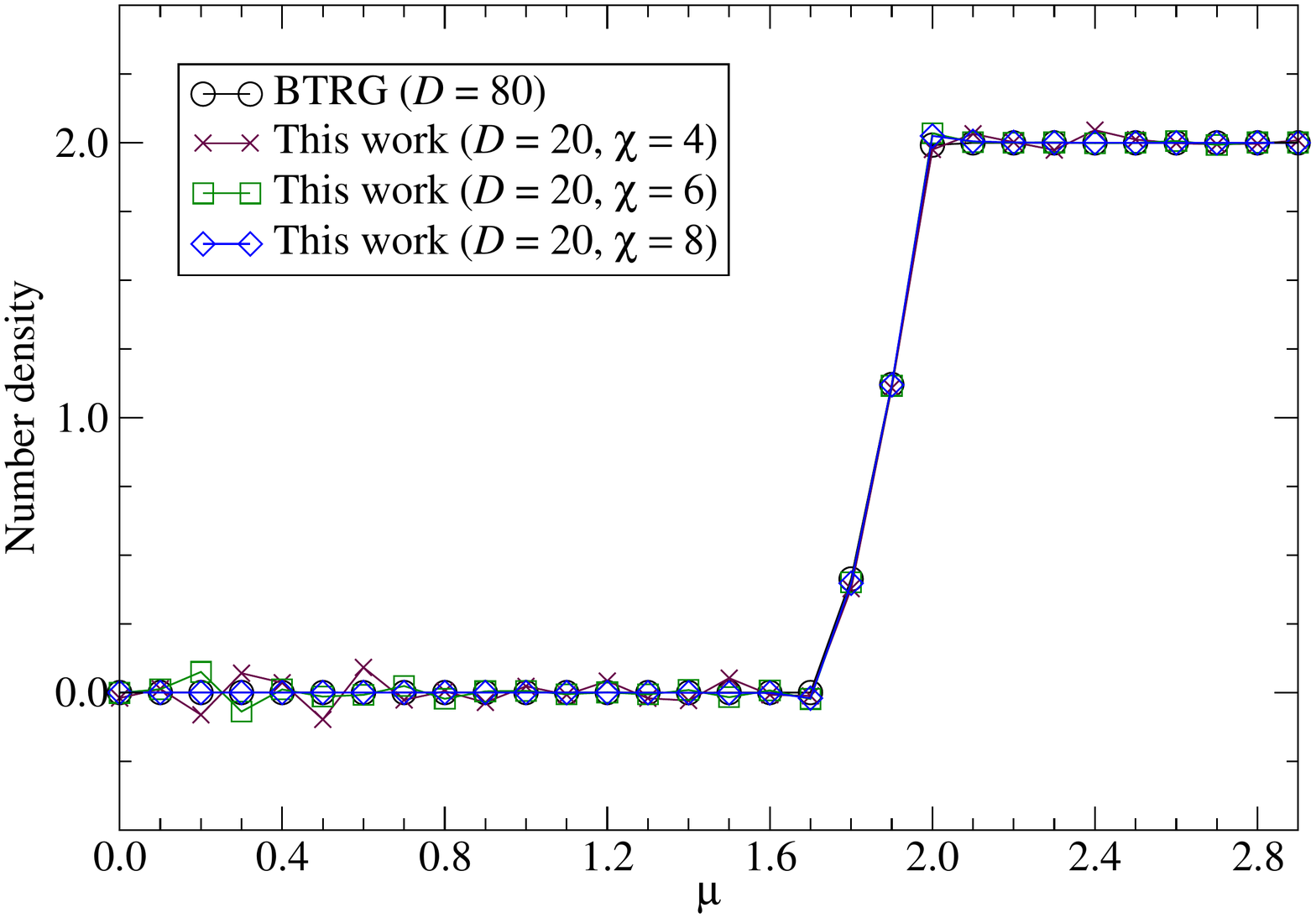}
	\caption{
	Number density as a function of chemical potential with $N_{f}=2$.
	Parameters and symbols are the same as those in Fig.~\ref{fig:nf2_g10}.
	}
	\label{fig:nf2_g10_number}
\end{figure}
\begin{figure}[htbp]
	\centering
	\includegraphics[width=0.9\hsize]{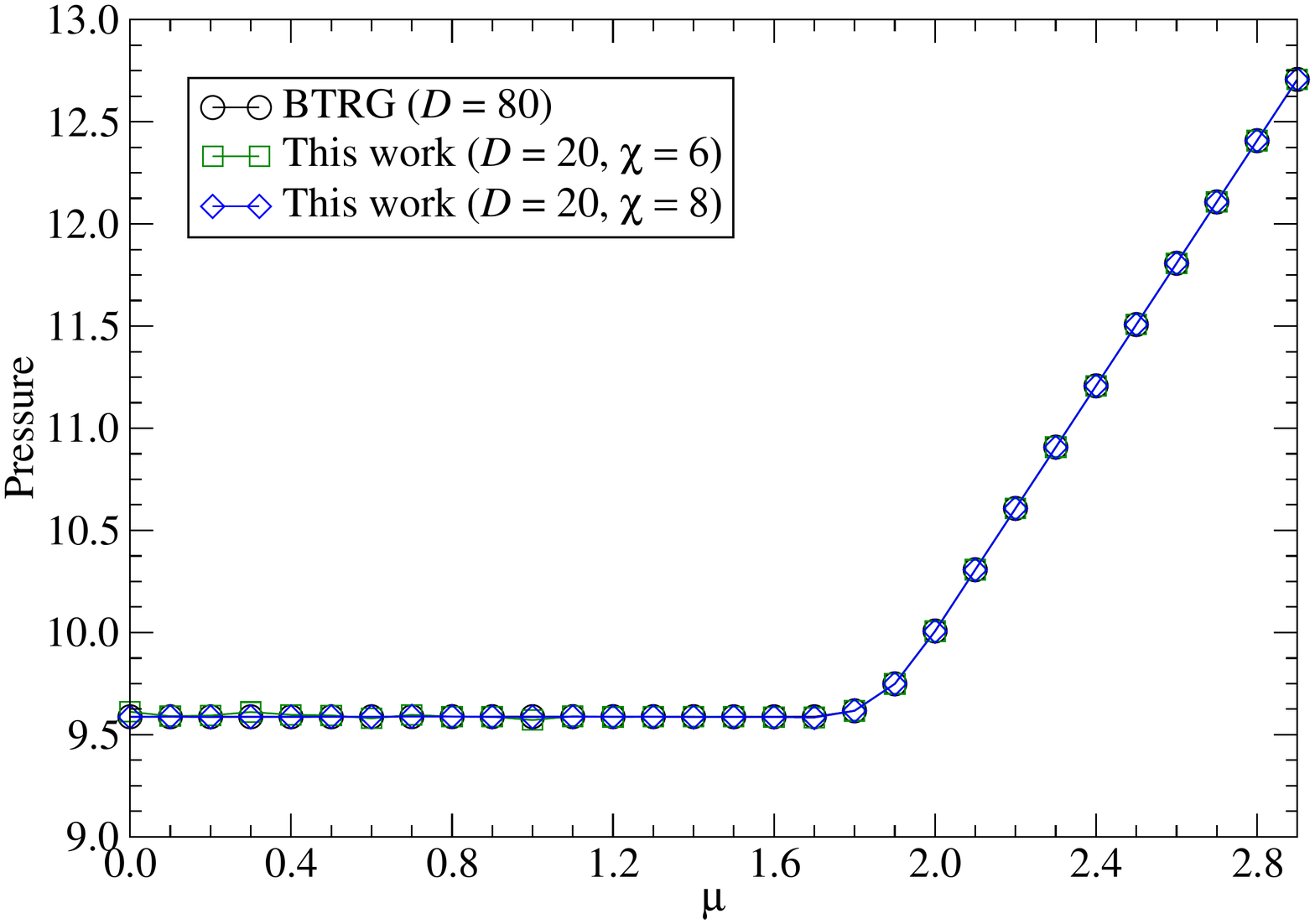}
	\caption{
	Pressure as a function of chemical potential with $N_{f}=3$, $g^2=10$, and $M=1$. 
	Square and diamond symbols correspond to computations with our algorithm setting $\chi=6,8$, respectively.
	The spacetime bond dimension is set as $D=20$ in all cases.
	Circles show the BTRG calculations, setting $D=80$.
	}
	\label{fig:nf3_g10}
\end{figure}
\begin{figure}[htbp]
	\centering
	\includegraphics[width=0.9\hsize]{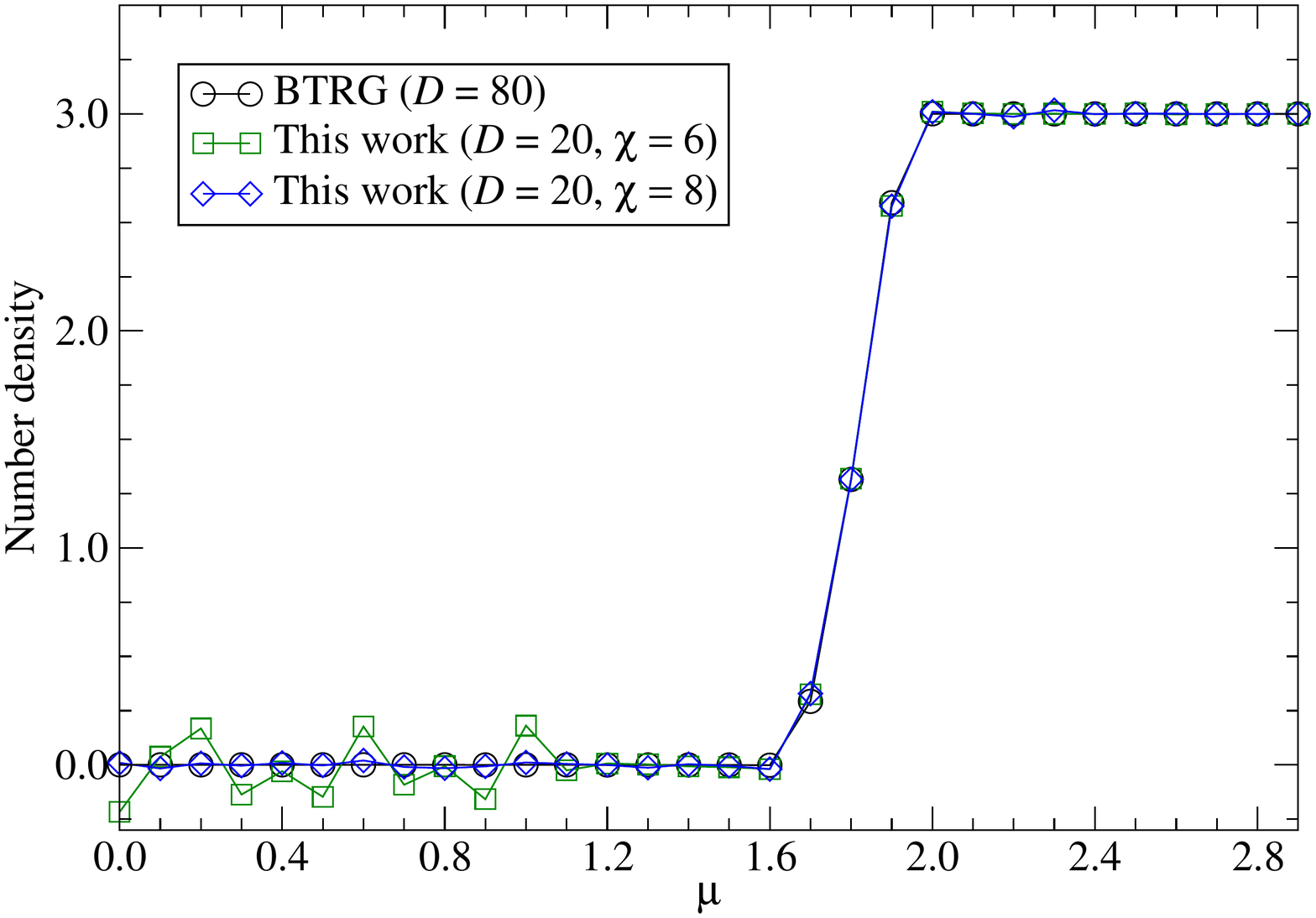}
	\caption{
	Number density as a function of chemical potential with $N_{f}=3$. 
	Parameters and symbols are the same as those in Fig.~\ref{fig:nf3_g10}.
	}
	\label{fig:nf3_g10_number}
\end{figure}

We now move on to the benchmark calculation with $g^{2}\neq0$ at finite density.
We evaluate pressure $P$ and number density $n$ on a lattice, whose size is $V=2^{20}$, as functions of chemical potential. 
In this work, the latter is calculated by the forward differentiation of the pressure such that
\begin{align}
	n=\frac{\partial P(\mu)}{\partial\mu}
	\approx
	\frac{P(\mu+\Delta\mu)-P(\mu)}{\Delta\mu}
\end{align}
with $\Delta\mu=0.1$.
We consider massive fermions with $M=1$ in order to observe a clear signal of the Silver Blaze phenomenon, which is a characteristic feature at finite density: bulk observables in the thermodynamic and zero-temperature limits do not depend on chemical potential up to a certain value $\mu_{\rm c}$~\cite{Cohen:2003kd}.
We choose $g^{2}=10$ as a representative.
We always set $D=20$ and vary the flavor bond dimension $\chi$.
Since we have observed the $\lo(10^{-7})$ relative error in the free field computation with $M=1$, we guess that $D=20$ is sufficiently large to suppress the finite-$D$ effect.
We also calculate the same quantities by the BTRG~\cite{PhysRevB.105.L060402} based on the two-dimensional Grassmann tensor network generated by Eq.~\eqref{eq:g_tensor} for validation.
\footnote{In the following, the hyperparameter $k$ in the BTRG algorithm is always set as $k=-0.5$, which is the optimal choice also in the case of lattice fermion, as shown in Ref.~\cite{Akiyama:2022pse}.}
Since the accuracy of the BTRG is higher than that of HOTRG at the same bond dimension~\cite{PhysRevB.105.L060402}, it would provide us with a good reference for massive fermion calculations.

Figures~\ref{fig:nf2_g10} and \ref{fig:nf2_g10_number} show pressure and number density with $N_{f}=2$ as functions of the chemical potential $\mu$. 
To see $\chi$-dependence, we have varied it as $\chi=4,6,8$.
Note that the GNW model with $N_{f}=2$ can be described as the two-layer Grassmann tensor network, whose bond dimension is equal to four.
The results obtained by our algorithm are consistent with those of the BTRG, where the Silver Blaze phenomenon is clearly observed.
Since we are considering the lattice Wilson fermion with $N_{f}=2$, the behavior of number density is also reasonable.
The endpoint of the Silver Blaze phenomenon seems around $\mu_{\rm c}\sim1.7$.
The calculations with $\chi=4, 6$ in Fig.~\ref{fig:nf2_g10_number} suggest that the finite-$\chi$ effect is more pronounced in the Silver Blaze regime.

We make a similar comparison between our algorithm and the BTRG in the case of $N_{f}=3$.
In Fig.~\ref{fig:nf3_g10}, we see that results obtained by both methods are consistent and capture the signal of the Silver Blaze phenomenon.
As shown in Fig.~\ref{fig:nf3_g10_number}, the maximal number of fermions per site should be $3$ because of $N_{f}=3$.
Again, we find that the finite-$\chi$ effect is more pronounced in the Silver Blaze regime.
However, the computation is stabilized with $\chi=8$ and the plateau of $n=0$ is restored.
The endpoint of the Silver Blaze phenomenon seems around $\mu_{\rm c}\sim1.6$ with $N_{f}=3$.

\section{Summary and outlook} 
\label{sec:summary}

Using the matrix product decomposition, we have shown that the lattice Gross-Neveu model with $N_{f}=2, 3$ can be described as the two- and three-layer Grassmann tensor network, whose bond dimension is four.
Based on this multi-layer representation, we have developed the coarse-graining procedure, where we convert two adjacent coefficient tensors into a canonical form along the virtual direction before we carry out the spacetime renormalization.
Thanks to the canonical form, we can obtain reduced density matrices easily and develop the coarse-graining procedure of the Grassmann tensor network via local operations with respect to each flavor.
With vanishing interaction, our algorithm automatically reduces the three-flavor computation into the single-flavor one.
This feature suggests that our algorithm show better performance, particularly in weak-coupling regime with lighter mass, compared with conventional TRG algorithms.
As a benchmark, we have calculated pressure and number density as functions of chemical potential on the lattice with $V=2^{20}$.
Although the finite-$\chi$ effect tends to be more pronounced in the Silver Blaze regime, the phenomena have been clearly observed with relatively small $\chi$ and $D$.

There are several future research directions.
If we could have the right-hand side of Eq.~\eqref{eq:vfc_nf3} analytically in advance, not via the SVD of the left-hand side of Eq.~\eqref{eq:vfc_nf3} numerically, the memory cost is reduced from $\lo(\e^{N_{f}})$ to $\lo(N_{f})$.
Since local Grassmann tensors are usually sparse, it can be expected that the geometry of the network can be rigorously deformed in other fermionic systems, not only the GNW model.
It seems very interesting to extend this study to four-dimensional lattice fermions with two and three flavors because they are necessary ingredients of the lattice quantum chromodynamics.
In addition, since our coarse-graining algorithm for $d$-dimensional lattice fermions is similar to the $(d+1)$-dimensional HOTRG, we can reduce the computational cost by considering the $(d+1)$-dimensional anisotropic TRG (ATRG)~\cite{Adachi:2019paf}.
Moreover, the coarse-graining of the $\Gamma$ tensor for each flavor (as shown in Eqs.~\eqref{eq:cg_x1}, \eqref{eq:cg_x2} and \eqref{eq:cg_x3}) can be done separately for each flavor, which is also compatible with parallel computation.
As another future work, we are also planning to investigate the Grassmann tensor network formulation for the spin-$S$ system with $S=1/2$ or $S=1$. We expect that the current coarse-graining procedure is promising to study such a system with the TRG approach.

\begin{acknowledgments}
We thank Tsuyoshi Okubo and Synge Todo for their valuable discussion.
The computation in this work has been done using the facilities of the Supercomputer Center, the Institute for Solid State Physics, the University of Tokyo.
We acknowledge the support from the Endowed Project for Quantum Software Research and Education, the University of Tokyo (\cite{qsw}). 
\end{acknowledgments}

\appendix

\section{Algorithmic details of coarse-graining procedure}
\label{app:algorithmic_details}

We demonstrate how to perform a single coarse-graining transformation.
We omit the iteration number $q$ as shown in Eq.~\eqref{eq:q} from all tensors.

\subsection{Conversion to the canonical form using flavor squeezers}
\label{subsec:flavor_cg}

\begin{figure}[htbp]
	\centering
	\includegraphics[width=1.0\hsize]{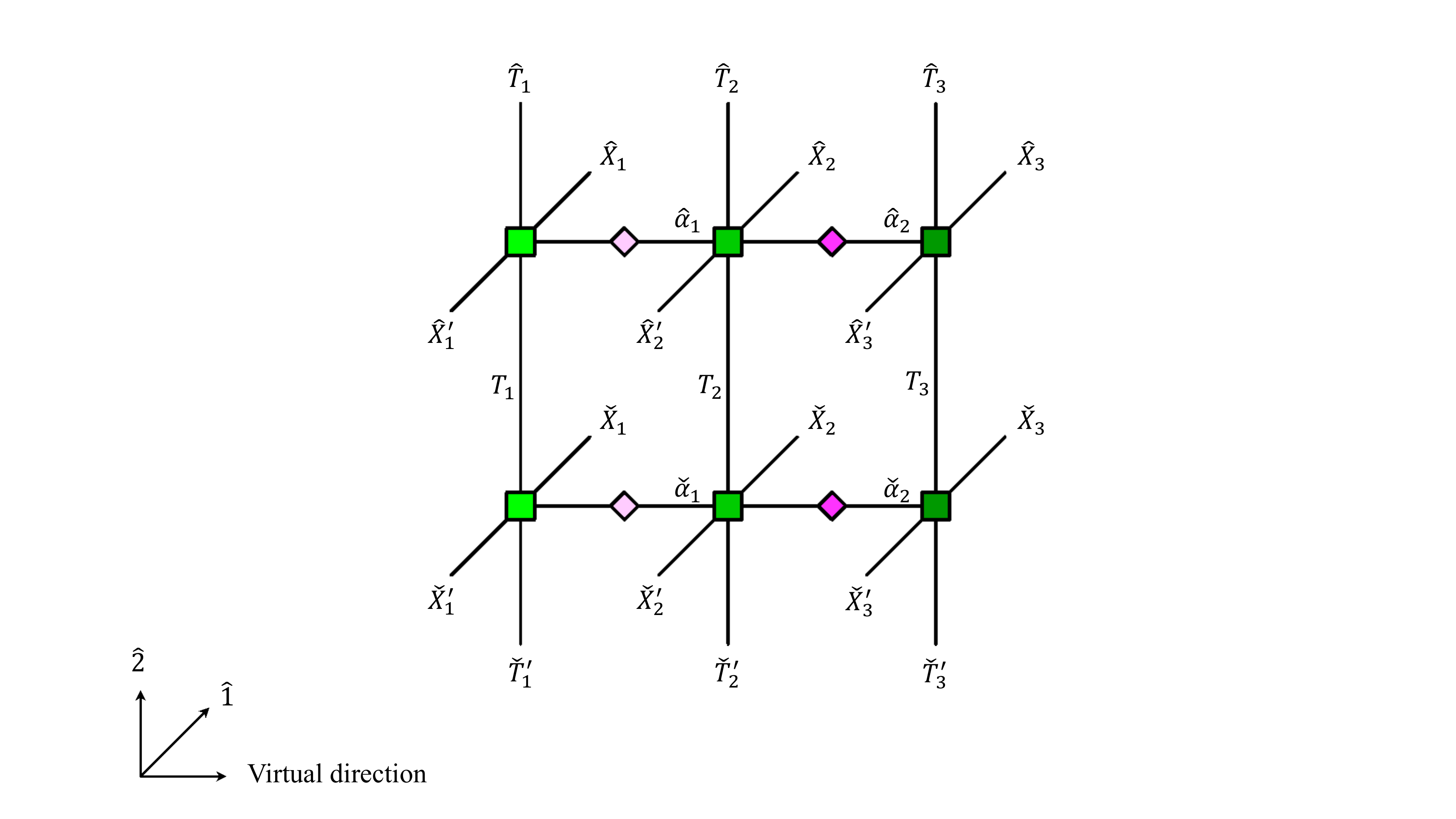}
	\caption{
	Illustration of temporally adjacent coefficient tensors.
	Square and diamond symbols are $\Gamma$'s  and $\sigma$'s, respectively, in Eqs.~\eqref{eq:hat} and \eqref{eq:check}.
	}
  	\label{fig:index}
\end{figure}

Let us rewrite Eq.~\eqref{eq:vfc_nf3} as
\begin{align}
\label{eq:init_mps}
	&T^{X_{1}T_{1}X'_{1}T'_{1}X_{2}T_{2}X'_{2}T'_{2}X_{3}T_{3}X'_{3}T'_{3}}
	\nonumber\\
	&=
	\sum_{\alpha_{1}=1}^{\chi}\sum_{\alpha_{2}=1}^{\chi}
	\Gamma^{X_{1}T_{1}X'_{1}T'_{1}}_{\alpha_{1}}
	\sigma_{\alpha_{1}}
	\Gamma^{X_{2}T_{2}X'_{2}T'_{2}}_{\alpha_{1}\alpha_{2}}
	\sigma_{\alpha_{2}}
	\Gamma^{X_{3}T_{3}X'_{3}T'_{3}}_{\alpha_{2}}
	.
\end{align}
\footnote{
When $q=0$, all the upper subscripts in Eq.~\eqref{eq:init_mps} are the two-bit strings defined via $X_{f}=x_{1}^{(f)}x_{2}^{(f)}$, $T_{f}=t_{1}^{(f)}t_{2}^{(f)}$, $X'_{f}=x_{1}^{(f)'}x_{2}^{(f)'}$, and $T'_{f}=t_{1}^{(f)'}t_{2}^{(f)'}$ with $f=1,2,3$.
}
We now introduce binary functions on these subscripts such that
\begin{widetext}
\begin{align}
\label{eq:binary_func}
	f_{I}(I)=
	\begin{cases}
		0~~~{\mathrm{if}}~I~{\mathrm{corresponds}~\mathrm{to}~\mathrm{the}~\mathrm{Grassmann}~\mathrm{even}~\mathrm{combination}} \\
		1~~~{\mathrm{if}}~I~{\mathrm{corresponds}~\mathrm{to}~\mathrm{the}~\mathrm{Grassmann}~\mathrm{odd}~\mathrm{combination}} 
	\end{cases}
	,
\end{align}
\end{widetext}
with $I=X_{f},T_{f},X'_{f},T'_{f}$ for $f=1,2,3$.
These binary functions are helpful to restore the Grassmann calculus just within coefficient tensors~\cite{Akiyama:2021nhe}. 
Let us focus on adjacent coefficient tensors along the temporal direction and set
\begin{align}
\label{eq:hat}
	&T^{\hat{X}_{1}\hat{T}_{1}\hat{X}'_{1}T_{1}\hat{X}_{2}\hat{T}_{2}\hat{X}'_{2}T_{2}\hat{X}_{3}\hat{T}_{3}\hat{X}'_{3}T_{3}}
	\nonumber\\
	&=
	\sum_{\hat{\alpha}_{1}=1}^{\chi}\sum_{\hat{\alpha}_{2}=1}^{\chi}
	\Gamma^{\hat{X}_{1}\hat{T}_{1}\hat{X}'_{1}T_{1}}_{\hat{\alpha}_{1}}
	\sigma_{\hat{\alpha}_{1}}
	\Gamma^{\hat{X}_{2}\hat{T}_{2}\hat{X}'_{2}T_{2}}_{\hat{\alpha}_{1}\hat{\alpha}_{2}}
	\sigma_{\hat{\alpha}_{2}}
	\Gamma^{\hat{X}_{3}\hat{T}_{3}\hat{X}'_{3}T_{3}}_{\hat{\alpha}_{2}}
\end{align}
be a coefficient tensor on a lattice site $n+\hat{2}$, and 
\begin{align}
\label{eq:check}
	&T^{\check{X}_{1}T_{1}\check{X}'_{1}\check{T}'_{1}\check{X}_{2}T_{2}\check{X}'_{2}\check{T}'_{2}\check{X}_{3}T_{3}\check{X}'_{3}\check{T}'_{3}}
	\nonumber\\
	&=
	\sum_{\check{\alpha}_{1}=1}^{\chi}\sum_{\check{\alpha}_{2}=1}^{\chi}
	\Gamma^{\check{X}_{1}T_{1}\check{X}'_{1}\check{T}'_{1}}_{\check{\alpha}_{1}}
	\sigma_{\check{\alpha}_{1}}
	\Gamma^{\check{X}_{2}T_{2}\check{X}'_{2}\check{T}'_{2}}_{\check{\alpha}_{1}\check{\alpha}_{2}}
	\sigma_{\check{\alpha}_{2}}
	\Gamma^{\check{X}_{3}T_{3}\check{X}'_{3}\check{T}'_{3}}_{\check{\alpha}_{2}}
\end{align}
on a site $n$ (Fig.~\ref{fig:index}).
Integrating out auxiliary Grassmann variables on the link $(n,n+\hat{2})$, which are shared by these two tensors, we have a new Grassmann tensor whose coefficient tensor is given by
\begin{widetext}
\begin{align}
\label{eq:combined_t}
	&(TT)^{\hat{X}_{1}\check{X}_{1}\hat{T}_{1}\hat{X}'_{1}\check{X}'_{1}\check{T}'_{1}\hat{X}_{2}\check{X}_{2}\hat{T}_{2}\hat{X}'_{2}\check{X}'_{2}\check{T}'_{2}\hat{X}_{3}\check{X}_{3}\hat{T}_{3}\hat{X}'_{3}\check{X}'_{3}\check{T}'_{3}}
	=
	\sum_{\hat{\alpha}_{1}=1}^{\chi}
	\sum_{\hat{\alpha}_{2}=1}^{\chi}
	\sum_{\check{\alpha}_{1}=1}^{\chi}
	\sum_{\check{\alpha}_{2}=1}^{\chi}
	\sum_{T_{1},T_{2},T_{3}}
	(-1)^{f_{T_{1}}(T_{1})+f_{T_{2}}(T_{2})+f_{T_{3}}(T_{3})}
	\nonumber\\
	&\times
	\Gamma^{\hat{X}_{1}\hat{T}_{1}\hat{X}'_{1}T_{1}}_{\hat{\alpha}_{1}}
	\sigma_{\hat{\alpha}_{1}}
	\Gamma^{\hat{X}_{2}\hat{T}_{2}\hat{X}'_{2}T_{2}}_{\hat{\alpha}_{1}\hat{\alpha}_{2}}
	\sigma_{\hat{\alpha}_{2}}
	\Gamma^{\hat{X}_{3}\hat{T}_{3}\hat{X}'_{3}T_{3}}_{\hat{\alpha}_{2}}
	\Gamma^{\check{X}_{1}T_{1}\check{X}'_{1}\check{T}'_{1}}_{\check{\alpha}_{1}}
	\sigma_{\check{\alpha}_{1}}
	\Gamma^{\check{X}_{2}T_{2}\check{X}'_{2}\check{T}'_{2}}_{\check{\alpha}_{1}\check{\alpha}_{2}}
	\sigma_{\check{\alpha}_{2}}
	\Gamma^{\check{X}_{3}T_{3}\check{X}'_{3}\check{T}'_{3}}_{\check{\alpha}_{2}}
	.
\end{align}
\end{widetext}
Note that $(-1)^{f_{T_{1}}(T_{1})+f_{T_{2}}(T_{2})+f_{T_{3}}(T_{3})}$ is resulting from the integration over auxiliary Grassmann variables on the link $(n,n+\hat{2})$.

\begin{figure*}[htbp]
	\centering
	\includegraphics[width=1.0\hsize]{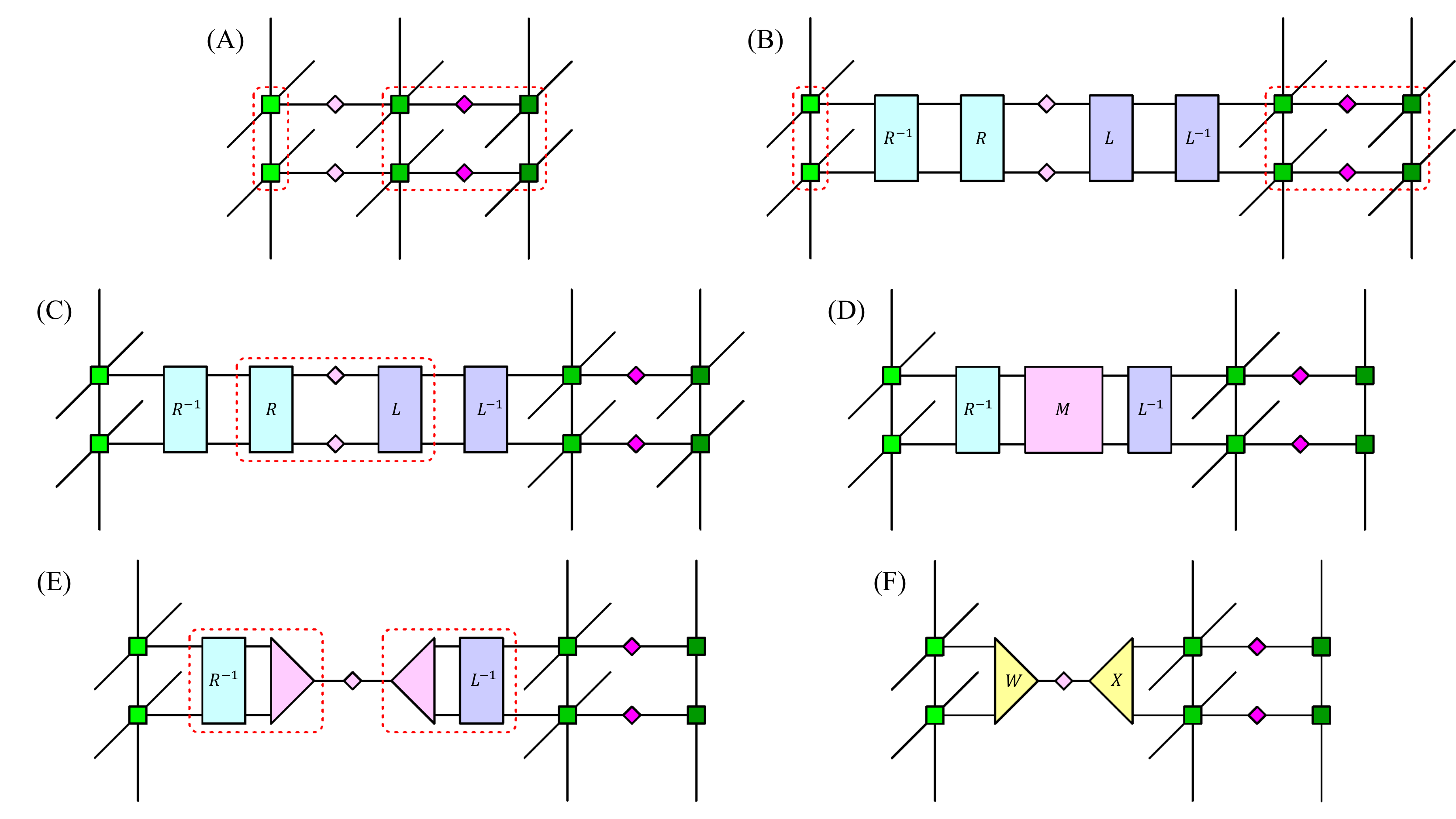}
	\caption{
	Illustration of how to derive flavor squeezers $W$ and $X$.
	(A) Diagrammatic representation of Eq.~\eqref{eq:combined_t}. Tensor contraction surrounded by a red dotted curve on the left shows $A$ in Eq.~\eqref{eq:a_tensor}. That surrounded by a red dotted curve on the right shows $B$ in Eq.~\eqref{eq:b_tensor}.
	(B) Inserting $R$($L$) and its inverse to the right(left) of $A$($B$). 
	(C) Tensor contraction in red dotted curve corresponds with the right-hand side of Eq.~\eqref{eq:def_m}.
	(D) Carrying out the contraction in (C), we have $M$ in the left-hand side of Eq.~\eqref{eq:def_m}.
	(E) Truncated SVD of $M$ with the flavor bond dimension $\chi$. Tensor contraction in a red dotted curve on the left defines the flavor squeezer $W$ in Eq.~\eqref{eq:def_w}. Similarly, tensor contraction in a red dotted curve on the right defines the flavor squeezer $X$ in Eq.~\eqref{eq:def_x}. 
	(F) $W$ and $X$ indirectly give us the SVD-based low-rank approximation of Eq.~\eqref{eq:combined_t}.
	}
  	\label{fig:cg_step_1}
\end{figure*}

\begin{figure*}[htbp]
	\centering
	\includegraphics[width=1.0\hsize]{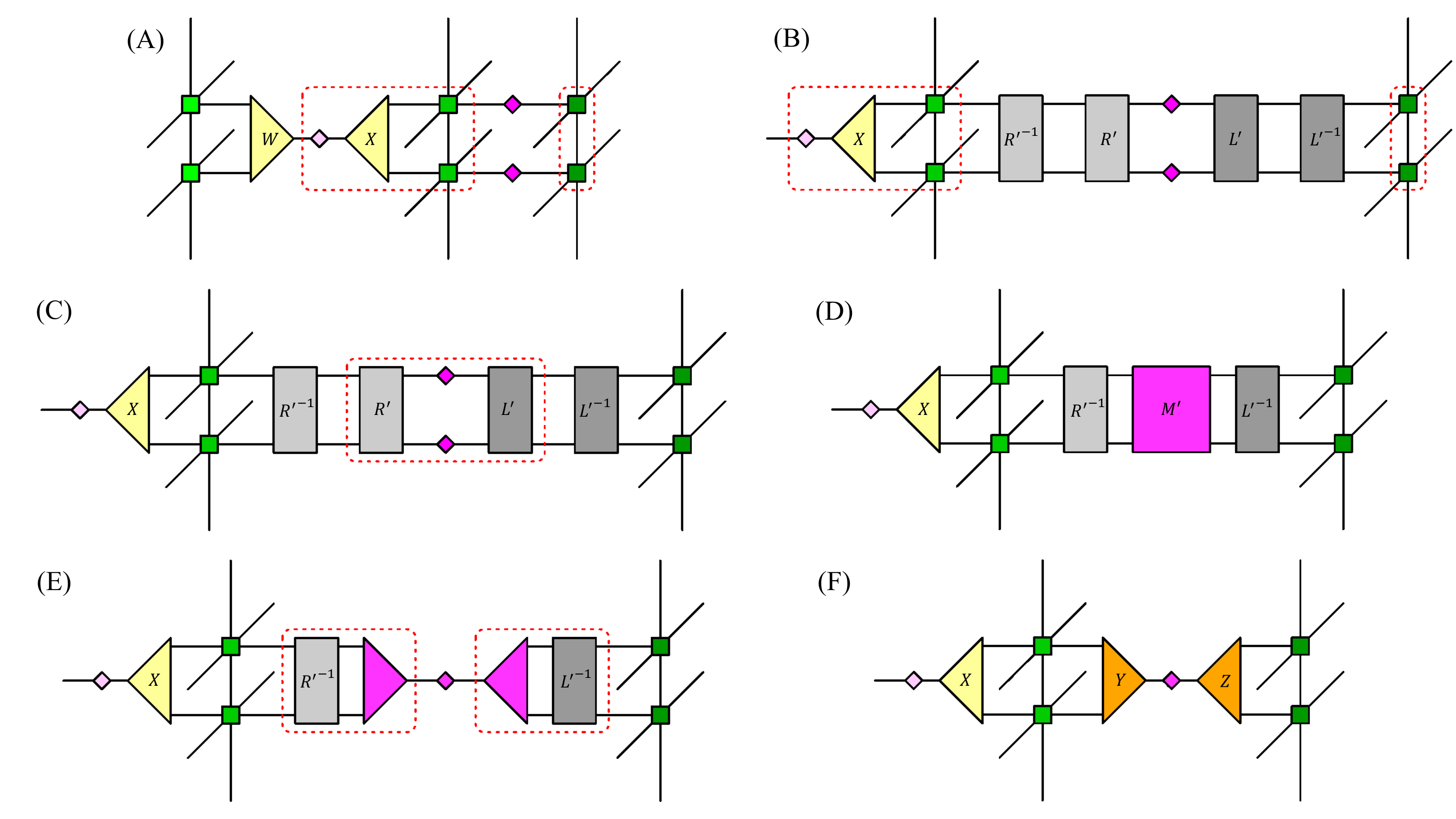}
	\caption{
	Illustration of how to derive flavor squeezers $Y$ and $Z$.
	(A) Tensor contraction surrounded by a red dotted curve on the left shows $C$ in Eq.~\eqref{eq:def_c}. That surrounded by a red dotted curve on the right shows $D$ in Eq.~\eqref{eq:def_d}.
	(B) Inserting $R'$($L'$) and its inverse to the right(left) of $C$($D$). 
	(C) Tensor contraction in a red dotted curve corresponds with the right-hand side of Eq.~\eqref{eq:def_m_prime}.
	(D) Carrying out the contraction in (C), we have $M'$ in the left-hand side of Eq.~\eqref{eq:def_m_prime}.
	(E) Truncated SVD of $M'$ with the flavor bond dimension $\chi$. Tensor contraction in a red dotted curve on the left defines the flavor squeezer $Y$ in Eq.~\eqref{eq:def_y}. Similarly, tensor contraction in a red dotted curve on the right defines the flavor squeezer $Z$ in Eq.~\eqref{eq:def_z}. 
	(F) $Y$ and $Z$ indirectly give us the SVD-based low-rank approximation of Eq.~\eqref{eq:cssd}.
	}
  	\label{fig:cg_step_2}
\end{figure*}

Let us convert Eq.~\eqref{eq:combined_t} into a canonical form along the virtual direction. 
A naive way is to directly carry out the SVD of Eq.~\eqref{eq:combined_t}, but here we consider an indirect way, which requires less memory space. Setting
\begin{align}
\label{eq:a_tensor}
	A^{\hat{X}_{1}\check{X}_{1}\hat{T}_{1}\hat{X}'_{1}\check{X}'_{1}\check{T}'_{1}}_{\hat{\alpha}_{1}\check{\alpha}_{1}}
	=
	\sum_{T_{1}}
	(-1)^{f_{T_{1}}(T_{1})}
	\Gamma^{\hat{X}_{1}\hat{T}_{1}\hat{X}'_{1}T_{1}}_{\hat{\alpha}_{1}}
	\Gamma^{\check{X}_{1}T_{1}\check{X}'_{1}\check{T}'_{1}}_{\check{\alpha}_{1}}
	,
\end{align}
\begin{align}
\label{eq:b_tensor}
	&B^{
	\hat{X}_{2}\check{X}_{2}\hat{T}_{2}\hat{X}'_{2}\check{X}'_{2}\check{T}'_{2}
	\hat{X}_{3}\check{X}_{3}\hat{T}_{3}\hat{X}'_{3}\check{X}'_{3}\check{T}'_{3}
	}_{\hat{\alpha}_{1}\check{\alpha}_{1}}
	\nonumber\\
	&=
	\sum_{\hat{\alpha}_{2},\check{\alpha}_{2}}
	\sum_{T_{2},T_{3}}
	(-1)^{f_{T_{2}}(T_{2})+f_{T_{3}}(T_{3})}
	\nonumber\\
	&\times
	\Gamma^{\hat{X}_{2}\hat{T}_{2}\hat{X}'_{2}T_{2}}_{\hat{\alpha}_{1}\hat{\alpha}_{2}}
	\sigma_{\hat{\alpha}_{2}}
	\Gamma^{\hat{X}_{3}\hat{T}_{3}\hat{X}'_{3}T_{3}}_{\hat{\alpha}_{2}}
	\Gamma^{\check{X}_{2}T_{2}\check{X}'_{2}\check{T}'_{2}}_{\check{\alpha}_{1}\check{\alpha}_{2}}
	\sigma_{\check{\alpha}_{2}}
	\Gamma^{\check{X}_{3}T_{3}\check{X}'_{3}\check{T}'_{3}}_{\check{\alpha}_{2}}
	,
\end{align}
we introduce an invertible matrix $R$ such that
\begin{align}
\label{eq:condition_r}
	\sum_{\hat{\alpha}_{1},\check{\alpha}_{1}}
	A^{\hat{X}_{1}\check{X}_{1}\hat{T}_{1}\hat{X}'_{1}\check{X}'_{1}\check{T}'_{1}}_{\hat{\alpha}_{1}\check{\alpha}_{1}}
	R^{-1}_{\hat{\alpha}_{1}\check{\alpha}_{1}\beta}
	=
	Q^{\hat{X}_{1}\check{X}_{1}\hat{T}_{1}\hat{X}'_{1}\check{X}'_{1}\check{T}'_{1}}_{\beta}
	,
\end{align}
where $Q$ satisfies the following condition,
\begin{align}
\label{eq:ortho_1}
	\sum_{\hat{X}_{1},\check{X}_{1},\hat{T}_{1},\hat{X}'_{1},\check{X}'_{1},\check{T}'_{1}}
	Q^{\hat{X}_{1}\check{X}_{1}\hat{T}_{1}\hat{X}'_{1}\check{X}'_{1}\check{T}'_{1}}_{\beta}
	{Q^{*}}^{\hat{X}_{1}\check{X}_{1}\hat{T}_{1}\hat{X}'_{1}\check{X}'_{1}\check{T}'_{1}}_{\beta'}
	=
	\delta_{\beta\beta'}
	.
\end{align}
To derive $R$, it is useful to construct a reduced density matrix from $A$,
\begin{align}
\label{eq:red_A}
	&\rho_{\hat{\alpha}_{1}\check{\alpha}_{1}\hat{\alpha}'_{1}\check{\alpha}'_{1}}
	\nonumber\\
	&=
	\sum_{\hat{X}_{1},\check{X}_{1},\hat{T}_{1},\hat{X}'_{1},\check{X}'_{1},\check{T}'_{1}}
	A^{\hat{X}_{1}\check{X}_{1}\hat{T}_{1}\hat{X}'_{1}\check{X}'_{1}\check{T}'_{1}}_{\hat{\alpha}_{1}\check{\alpha}_{1}}
	{A^{*}}^{\hat{X}_{1}\check{X}_{1}\hat{T}_{1}\hat{X}'_{1}\check{X}'_{1}\check{T}'_{1}}_{\hat{\alpha}'_{1}\check{\alpha}'_{1}}
	,
\end{align}
whose eigenvalue decomposition (EVD) provides us with $R$ via
\begin{align}
\label{eq:norm_tensor}
	R_{\hat{\alpha}_{1}\check{\alpha}_{1}\beta}
	=
	U_{\hat{\alpha}_{1}\check{\alpha}_{1}\beta}
	\sqrt{\lambda_{\beta}}
	,
\end{align}
where $U$ diagonalizes the reduced density matrix in Eq.~\eqref{eq:red_A} and $\lambda_{\beta}$ is the eigenvalue. Similarly, we can find an invertible matrix $L$ such that
\begin{align}
\label{eq:ortho_2}
	\sum_{\hat{\alpha}_{1},\check{\alpha}_{1}}
	L^{-1}_{\gamma\hat{\alpha}_{1}\check{\alpha}_{1}}
	B^{
	\hat{X}_{2}\cdots\check{T}'_{3}
	}_{\hat{\alpha}_{1}\check{\alpha}_{1}}
	=
	Q^{
	\hat{X}_{2}\cdots\check{T}'_{3}
	}_{\gamma}
	,
\end{align}
with the following orthogonality condition,
\begin{align}
\label{eq:ortho_23}
	\sum_{\hat{X}_{2},\cdots,\check{T}'_{3}}
	Q^{
	\hat{X}_{2}\cdots\check{T}'_{3}
	}_{\gamma}
	{Q^{*}}^{
	\hat{X}_{2}\cdots\check{T}'_{3}
	}_{\gamma'}
	=
	\delta_{\gamma\gamma'}
	.
\end{align}
Again, the reduced density matrix constructed from $B$ provides us with a possible choice of $L$. Defining
\begin{align}
\label{eq:def_m}
	M_{\beta\gamma}
	= 
	\sum_{\hat{\alpha}_{1},\check{\alpha}_{1}}
	R_{\hat{\alpha}_{1}\check{\alpha}_{1}\beta}
	\sigma_{\hat{\alpha}_{1}}
	\sigma_{\check{\alpha}_{1}}
	L_{\hat{\alpha}_{1}\check{\alpha}_{1}\gamma}
	,
\end{align}
whose low-rank approximation is
\begin{align}
\label{eq:svd_m}
	M_{\beta\gamma}
	\approx
	\sum_{\alpha_{1}=1}^{\chi}
	U_{\beta\alpha_{1}}
	\sigma_{\alpha_{1}}
	V^{*}_{\gamma\alpha_{1}}
	,
\end{align}
we can introduce three-leg tensors,
\begin{align}
\label{eq:def_w}
	W_{\hat{\alpha}_{1}\check{\alpha}_{1}\alpha_{1}}
	=
	\sum_{\beta}
	R^{-1}_{\hat{\alpha}_{1}\check{\alpha}_{1}\beta}
	U_{\beta\alpha_{1}}
	,
\end{align}
\begin{align}
\label{eq:def_x}
	X_{\alpha_{1}\hat{\alpha}_{1}\check{\alpha}_{1}}
	=
	\sum_{\gamma}
	V^{*}_{\gamma\alpha_{1}}
	L^{-1}_{\gamma\hat{\alpha}_{1}\check{\alpha}_{1}}
	.
\end{align}
Using $W$ and $X$, we can indirectly obtain
\begin{align}
\label{eq:1st_svd_form}
	&(TT)^{
	\hat{X}_{1}\cdots\check{T}'_{1}
	\hat{X}_{2}\cdots\check{T}'_{2}
	\hat{X}_{3}\cdots\check{T}'_{3}
	}
	\approx
	\sum_{\alpha_{1}=1}^{\chi}
	\sum_{\beta,\gamma}
	\nonumber\\
	&
	Q^{\hat{X}_{1}\check{X}_{1}\hat{T}_{1}\hat{X}'_{1}\check{X}'_{1}\check{T}'_{1}}_{\beta}
	U_{\beta\alpha_{1}}
	\sigma_{\alpha_{1}}
	V^{*}_{\gamma\alpha_{1}}
	Q^{
	\hat{X}_{2}\check{X}_{2}\hat{T}_{2}\hat{X}'_{2}\check{X}'_{2}\check{T}'_{2}
	\hat{X}_{3}\check{X}_{3}\hat{T}_{3}\hat{X}'_{3}\check{X}'_{3}\check{T}'_{3}
	}_{\gamma}
	,
\end{align}
which is the SVD of Eq.~\eqref{eq:combined_t}.
We refer to these three-leg tensors as flavor squeezers. 
The above procedure to derive $W$ and $X$ are diagrammatically summarized in Fig.~\ref{fig:cg_step_1}. \footnote{So far, $R$ and $L$ have been written as if they are of three legs. However, one can also regard them as four-leg tensors and Fig.~\ref{fig:cg_step_1} is based on the later description.}

Next, setting
\begin{align}
\label{eq:def_c}
	&C^{\hat{X}_{2}\check{X}_{2}\hat{T}_{2}\hat{X}'_{2}\check{X}'_{2}\check{T}'_{2}}_{\alpha_{1}\hat{\alpha}_{2}\check{\alpha}_{2}}
	\nonumber\\
	&=
	\sum_{\hat{\alpha}_{1},\check{\alpha}_{1}}
	\sum_{T_{2}}
	(-1)^{f_{T_{2}}(T_{2})}
	\sigma_{\alpha_{1}}
	X_{\alpha_{1}\hat{\alpha}_{1}\check{\alpha}_{1}}
	\Gamma^{\hat{X}_{2}\hat{T}_{2}\hat{X}'_{2}T_{2}}_{\hat{\alpha}_{1}\hat{\alpha}_{2}}
	\Gamma^{\check{X}_{2}T_{2}\check{X}'_{2}\check{T}'_{2}}_{\check{\alpha}_{1}\check{\alpha}_{2}}
	,
\end{align}
\begin{align}
\label{eq:def_d}
	D^{\hat{X}_{3}\check{X}_{3}\hat{T}_{3}\hat{X}'_{3}\check{X}'_{3}\check{T}'_{3}}_{\hat{\alpha}_{2}\check{\alpha}_{2}}
	=
	\sum_{T_{3}}
	(-1)^{f_{T_{3}}(T_{3})}
	\Gamma^{\hat{X}_{3}\hat{T}_{3}\hat{X}'_{3}T_{3}}_{\hat{\alpha}_{2}}
	\Gamma^{\check{X}_{3}T_{3}\check{X}'_{3}\check{T}'_{3}}_{\check{\alpha}_{2}}
	,
\end{align}
we consider invertible matrices $R'$ and $L'$ such that
\begin{align}
\label{eq:r_prime}
	\sum_{\hat{\alpha}_{2},\check{\alpha}_{2}}
	C^{\hat{X}_{2}\check{X}_{2}\hat{T}_{2}\hat{X}'_{2}\check{X}'_{2}\check{T}'_{2}}_{\alpha_{1}\hat{\alpha}_{2}\check{\alpha}_{2}}
	{R'}^{-1}_{\hat{\alpha}_{2}\check{\alpha}_{2}\beta}
	=
	Q^{\hat{X}_{2}\check{X}_{2}\hat{T}_{2}\hat{X}'_{2}\check{X}'_{2}\check{T}'_{2}}_{\alpha_{1}\beta}
	,
\end{align}
\begin{align}
\label{eq:l_prime}
	\sum_{\hat{\alpha}_{2},\check{\alpha}_{2}}
	{L'}^{-1}_{\gamma\hat{\alpha}_{2}\check{\alpha}_{2}}
	D^{\hat{X}_{3}\check{X}_{3}\hat{T}_{3}\hat{X}'_{3}\check{X}'_{3}\check{T}'_{3}}_{\hat{\alpha}_{2}\check{\alpha}_{2}}
	=
	Q^{\hat{X}_{3}\check{X}_{3}\hat{T}_{3}\hat{X}'_{3}\check{X}'_{3}\check{T}'_{3}}_{\gamma}
	,
\end{align}
with orthogonality conditions
\begin{align}
\label{eq:ortho_2}
	\sum_{\alpha_{1},\hat{X}_{2},\check{X}_{2},\hat{T}_{2},\hat{X}'_{2},\check{X}'_{2},\check{T}'_{2}}
	Q^{\hat{X}_{2}\check{X}_{2}\hat{T}_{2}\hat{X}'_{2}\check{X}'_{2}\check{T}'_{2}}_{\alpha_{1}\beta}
	{Q^{*}}^{\hat{X}_{2}\check{X}_{2}\hat{T}_{2}\hat{X}'_{2}\check{X}'_{2}\check{T}'_{2}}_{\alpha_{1}\beta'}
	=
	\delta_{\beta\beta'}
	,
\end{align}
\begin{align}
\label{eq:ortho_3}
	\sum_{\hat{X}_{3},\check{X}_{3},\hat{T}_{3},\hat{X}'_{3},\check{X}'_{3},\check{T}'_{3}}
	Q^{\hat{X}_{3}\check{X}_{3}\hat{T}_{3}\hat{X}'_{3}\check{X}'_{3}\check{T}'_{3}}_{\gamma}
	{Q^{*}}^{\hat{X}_{3}\check{X}_{3}\hat{T}_{3}\hat{X}'_{3}\check{X}'_{3}\check{T}'_{3}}_{\gamma'}
	=
	\delta_{\gamma\gamma'}
	.
\end{align}
Constructing a reduced density matrix from $C$ ($D$), one can find $R'$ ($L'$) in the same way with Eq.~\eqref{eq:norm_tensor} satisfying Eqs.~\eqref{eq:r_prime} and \eqref{eq:ortho_2} (Eqs.~\eqref{eq:l_prime} and \eqref{eq:ortho_3}). 
Introducing
\begin{align}
\label{eq:def_m_prime}
	M'_{\beta\gamma}
	= 
	\sum_{\hat{\alpha}_{2},\check{\alpha}_{2}}
	R'_{\hat{\alpha}_{2}\check{\alpha}_{2}\beta}
	\sigma_{\hat{\alpha}_{2}}
	\sigma_{\check{\alpha}_{2}}
	L'_{\hat{\alpha}_{2}\check{\alpha}_{2}\gamma}
	,
\end{align}
whose low-rank approximation is
\begin{align}
\label{eq:svd_m_prime}
	M'_{\beta\gamma}
	\approx
	\sum_{\alpha_{2}=1}^{\chi}
	U_{\beta\alpha_{2}}
	\sigma_{\alpha_{2}}
	V^{*}_{\gamma\alpha_{2}}
	,
\end{align}
we can derive flavor squeezers,
\begin{align}
\label{eq:def_y}
	Y_{\hat{\alpha}_{2}\check{\alpha}_{2}\alpha_{2}}
	=
	\sum_{\beta}
	{R'}^{-1}_{\hat{\alpha}_{2}\check{\alpha}_{2}\beta}
	U_{\beta\alpha_{2}}
	,
\end{align}
\begin{align}
\label{eq:def_z}
	Z_{\alpha_{2}\hat{\alpha}_{2}\check{\alpha}_{2}}
	=
	\sum_{\gamma}
	V^{*}_{\gamma\alpha_{2}}
	{L'}^{-1}_{\gamma\hat{\alpha}_{2}\check{\alpha}_{2}}
	.
\end{align}
Note that these flavor squeezers $Y$ and $Z$ enable us to indirectly carry out the SVD of
\begin{align}
\label{eq:cssd}
	\sum_{\hat{\alpha}_{2},\check{\alpha}_{2}}
	C^{\hat{X}_{2}\check{X}_{2}\hat{T}_{2}\hat{X}'_{2}\check{X}'_{2}\check{T}'_{2}}_{\alpha_{1}\hat{\alpha}_{2}\check{\alpha}_{2}}
	\sigma_{\hat{\alpha}_{2}}
	\sigma_{\check{\alpha}_{2}}
	D^{\hat{X}_{3}\check{X}_{3}\hat{T}_{3}\hat{X}'_{3}\check{X}'_{3}\check{T}'_{3}}_{\hat{\alpha}_{2}\check{\alpha}_{2}}
	,
\end{align}
whose singular value is $\sigma_{\alpha_{2}}$ in Eq.~\eqref{eq:svd_m_prime}. Figure~\ref{fig:cg_step_2} summarizes how to derive $Y$ and $Z$.

\subsection{Spacetime coarse-graining}
\label{subsec:sp_cg}

\begin{figure*}[htbp]
	\centering
	\includegraphics[width=1.0\hsize]{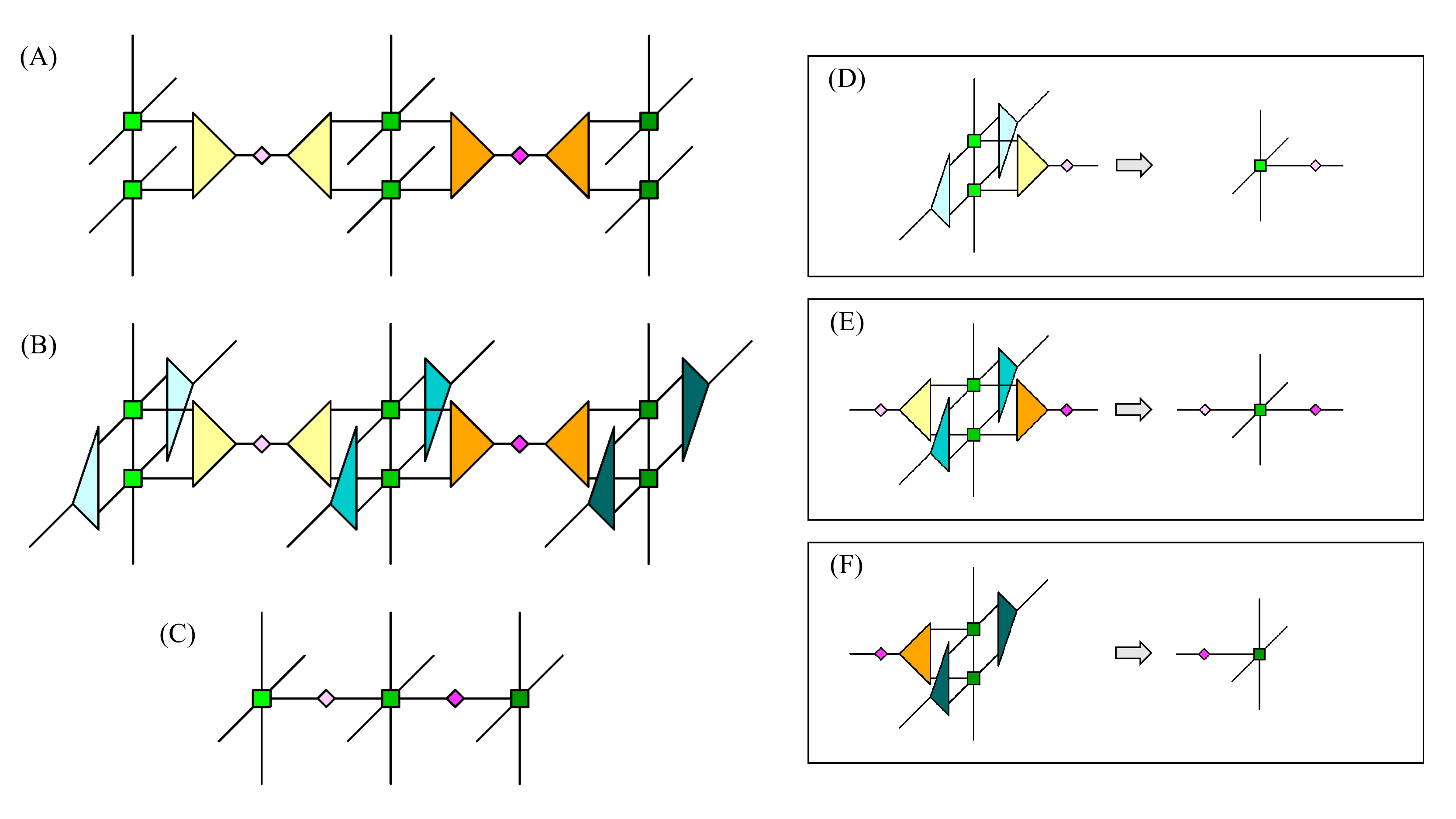}
	\caption{
	Schematic illustration of the spacetime coarse-graining along the temporal direction. 
	(A) Illustration of Eq.~\eqref{eq:new_vidal}.
	(B) Spacetime squeezers are inserted.
	(C) Tensor contraction provides us with a coarse-grained tensor, whose form is the same as the right-hand side of Eq.~\eqref{eq:init_mps}.
	(D) Diagram for Eq.~\eqref{eq:cg_x1}.
	(E) Diagram for Eq.~\eqref{eq:cg_x2}.
	(F) Diagram for Eq.~\eqref{eq:cg_x3}.
	}
  	\label{fig:cg_step_3}
\end{figure*}

Using flavor squeezers, $W$, $X$, $Y$, and $Z$ derived previously, we can express Eq.~\eqref{eq:combined_t} in the canonical form of
\begin{widetext}
\begin{align}
\label{eq:new_vidal}
	&(TT)^{
	\hat{X}_{1}\cdots\check{T}'_{1}
	\hat{X}_{2}\cdots\check{T}'_{2}
	\hat{X}_{3}\cdots\check{T}'_{3}
	}
	\approx
	\sum_{\alpha_{1}=1}^{\chi}
	\sum_{\alpha_{2}=1}^{\chi}
	\Gamma^{\hat{X}_{1}\check{X}_{1}\hat{T}_{1}\hat{X}'_{1}\check{X}'_{1}\check{T}'_{1}}_{\alpha_{1}}
	\sigma_{\alpha_{1}}
	\Gamma^{\hat{X}_{2}\check{X}_{2}\hat{T}_{2}\hat{X}'_{2}\check{X}'_{2}\check{T}'_{2}}_{\alpha_{1}\alpha_{2}}
	\sigma_{\alpha_{2}}
	\Gamma^{\hat{X}_{3}\check{X}_{3}\hat{T}_{3}\hat{X}'_{3}\check{X}'_{3}\check{T}'_{3}}_{\alpha_{2}}
	,
\end{align}
\end{widetext}
where $\sigma_{\alpha_{1}}$, $\sigma_{\alpha_{2}}$ are given in Eqs.~\eqref{eq:svd_m}, \eqref{eq:svd_m_prime}, respectively, and
\begin{align}
	\Gamma^{\hat{X}_{1}\check{X}_{1}\hat{T}_{1}\hat{X}'_{1}\check{X}'_{1}\check{T}'_{1}}_{\alpha_{1}}
	=
	\sum_{\hat{\alpha}_{1},\check{\alpha}_{1}}
	A^{\hat{X}_{1}\check{X}_{1}\hat{T}_{1}\hat{X}'_{1}\check{X}'_{1}\check{T}'_{1}}_{\hat{\alpha}_{1}\check{\alpha}_{1}}
	W_{\hat{\alpha}_{1}\check{\alpha}_{1}\alpha_{1}}
	,
\end{align}
\begin{align}
	\sigma_{\alpha_{1}}
	\Gamma^{\hat{X}_{2}\check{X}_{2}\hat{T}_{2}\hat{X}'_{2}\check{X}'_{2}\check{T}'_{2}}_{\alpha_{1}\alpha_{2}}
	=
	%\sum_{\hat{\alpha}_{1},\check{\alpha}_{1}}
	\sum_{\hat{\alpha}_{2},\check{\alpha}_{2}}
	C^{\hat{X}_{2}\check{X}_{2}\hat{T}_{2}\hat{X}'_{2}\check{X}'_{2}\check{T}'_{2}}_{\alpha_{1}\hat{\alpha}_{2}\check{\alpha}_{2}}
	Y_{\hat{\alpha}_{2}\check{\alpha}_{2}\alpha_{2}}
	,
\end{align}
\begin{align}
	\Gamma^{\hat{X}_{3}\check{X}_{3}\hat{T}_{3}\hat{X}'_{3}\check{X}'_{3}\check{T}'_{3}}_{\alpha_{2}}
	=
	\sum_{\hat{\alpha}_{2},\check{\alpha}_{2}}
	Z_{\alpha_{2}\hat{\alpha}_{2}\check{\alpha}_{2}}
	D^{\hat{X}_{3}\check{X}_{3}\hat{T}_{3}\hat{X}'_{3}\check{X}'_{3}\check{T}'_{3}}_{\hat{\alpha}_{2}\check{\alpha}_{2}}
	.
\end{align}
Note that the flavor squeezer $X$ has been employed to define $C$ as in Eq.~\eqref{eq:def_c}. 

Thanks to orthogonality conditions in Eqs.~\eqref{eq:ortho_1}, \eqref{eq:ortho_23}, \eqref{eq:ortho_2} and \eqref{eq:ortho_3}, we are allowed to carry out the HOTRG-like spacetime coarse-graining ``locally" with respect to each flavor segment. 
For example, let us consider the following density matrix,
\begin{align}
\label{eq:red_density_x1}
	&\rho^{\hat{X}_{1}\check{X}_{1}\tilde{\hat{X}}_{1}\tilde{\check{X}}_{1}}
	=
	\sum_{\hat{T}_{1},\hat{X}'_{1},\check{X}'_{1},\check{T}'_{1}}
	\sum_{\hat{X}_{2},\cdots,\check{T}'_{2}}
	\sum_{\hat{X}_{3},\cdots,\check{T}'_{3}}
	\nonumber\\
	&
	(TT)^{
	\hat{X}_{1}\check{X}_{1}\cdots\check{T}'_{1}
	\hat{X}_{2}\cdots\check{T}'_{2}
	\hat{X}_{3}\cdots\check{T}'_{3}
	}
	{(TT)^*}^{
	\tilde{\hat{X}}_{1}\tilde{\check{X}}_{1}\cdots\check{T}'_{1}
	\hat{X}_{2}\cdots\check{T}'_{2}
	\hat{X}_{3}\cdots\check{T}'_{3}
	}
	.
\end{align}
Using the orthogonality condition of Eq.~\eqref{eq:ortho_23} and the approximation of $(TT)$ in Eq.~\eqref{eq:1st_svd_form}, one can calculate $\rho^{\hat{X}_{1}\check{X}_{1}\tilde{\hat{X}}_{1}\tilde{\check{X}}_{1}}$ as
\begin{align}
	&\rho^{\hat{X}_{1}\check{X}_{1}\tilde{\hat{X}}_{1}\tilde{\check{X}}_{1}}
	\nonumber\\
	&=
	\sum_{\hat{T}_{1},\hat{X}'_{1},\check{X}'_{1},\check{T}'_{1}}
	\sum_{\alpha_{1}}
	\Gamma^{\hat{X}_{1}\check{X}_{1}\hat{T}_{1}\hat{X}'_{1}\check{X}'_{1}\check{T}'_{1}}_{\alpha_{1}}
	\sigma_{\alpha_{1}}
	{\Gamma^{*}}^{\tilde{\hat{X}}_{1}\tilde{\check{X}}_{1}\hat{T}_{1}\hat{X}'_{1}\check{X}'_{1}\check{T}'_{1}}_{\alpha_{1}}
	\sigma_{\alpha_{1}}
	.
\end{align}
One can define $\rho^{\hat{X}_{2}\check{X}_{2}\tilde{\hat{X}}_{2}\tilde{\check{X}}_{2}}$ and $\rho^{\hat{X}_{3}\check{X}_{3}\tilde{\hat{X}}_{3}\tilde{\check{X}}_{3}}$ in the same way with Eq.~\eqref{eq:red_density_x1}.
Using orthogonality conditions of Eqs.~\eqref{eq:ortho_2}, \eqref{eq:ortho_3} and the approximation of $(TT)$ in Eq.~\eqref{eq:new_vidal}, one obtains these reduced density matrices via
\begin{align}
	&\rho^{\hat{X}_{2}\check{X}_{2}\tilde{\hat{X}}_{2}\tilde{\check{X}}_{2}}
	=
	\sum_{\hat{T}_{2},\hat{X}'_{2},\check{X}'_{2},\check{T}'_{2}}
	\sum_{\alpha_{1},\alpha_{2}}
	\nonumber\\
	&
	\sigma_{\alpha_{1}}
	\Gamma^{\hat{X}_{2}\check{X}_{2}\hat{T}_{2}\hat{X}'_{2}\check{X}'_{2}\check{T}'_{2}}_{\alpha_{1}\alpha_{2}}
	\sigma_{\alpha_{2}}
	\sigma_{\alpha_{1}}
	{\Gamma^{*}}^{\tilde{\hat{X}}_{2}\tilde{\check{X}}_{2}\hat{T}_{2}\hat{X}'_{2}\check{X}'_{2}\check{T}'_{2}}_{\alpha_{1}\alpha_{2}}
	\sigma_{\alpha_{2}}
	,
\end{align}
\begin{align}
	&\rho^{\hat{X}_{3}\check{X}_{3}\tilde{\hat{X}}_{3}\tilde{\check{X}}_{3}}
	=
	\sum_{\hat{T}_{3},\hat{X}'_{3},\check{X}'_{3},\check{T}'_{3}}
	\sum_{\alpha_{2}}
	\nonumber\\
	&
	\sigma_{\alpha_{2}}
	\Gamma^{\hat{X}_{3}\check{X}_{3}\hat{T}_{3}\hat{X}'_{3}\check{X}'_{3}\check{T}'_{3}}_{\alpha_{2}}
	\sigma_{\alpha_{2}}
	{\Gamma^{*}}^{\tilde{\hat{X}}_{3}\tilde{\check{X}}_{3}\hat{T}_{3}\hat{X}'_{3}\check{X}'_{3}\check{T}'_{3}}_{\alpha_{2}}
	.
\end{align}
A similar discussion provides us with reduced density matrices such that
\begin{align}
	&\rho^{\hat{X}'_{1}\check{X}'_{1}\tilde{\hat{X}}'_{1}\tilde{\check{X}}'_{1}}
	=
	\sum_{\hat{X}_{1},\check{X}_{1},\hat{T}_{1},\check{T}'_{1}}
	\sum_{\alpha_{1}}
	\nonumber\\
	&
	\Gamma^{\hat{X}_{1}\check{X}_{1}\hat{T}_{1}\hat{X}'_{1}\check{X}'_{1}\check{T}'_{1}}_{\alpha_{1}}
	\sigma_{\alpha_{1}}
	{\Gamma^{*}}^{\hat{X}_{1}\check{X}_{1}\hat{T}_{1}\tilde{\hat{X}}'_{1}\tilde{\check{X}}'_{1}\check{T}'_{1}}_{\alpha_{1}}
	\sigma_{\alpha_{1}}
	,
\end{align}
\begin{align}
	&\rho^{\hat{X}'_{2}\check{X}'_{2}\tilde{\hat{X}}'_{2}\tilde{\check{X}}'_{2}}
	=
	\sum_{\hat{X}_{2},\check{X}_{2},\hat{T}_{2},\check{T}'_{2}}
	\sum_{\alpha_{1},\alpha_{2}}
	\nonumber\\
	&
	\sigma_{\alpha_{1}}
	\Gamma^{\hat{X}_{2}\check{X}_{2}\hat{T}_{2}\hat{X}'_{2}\check{X}'_{2}\check{T}'_{2}}_{\alpha_{1}\alpha_{2}}
	\sigma_{\alpha_{2}}
	\sigma_{\alpha_{1}}
	{\Gamma^{*}}^{\hat{X}_{2}\check{X}_{2}\hat{T}_{2}\tilde{\hat{X}}'_{2}\tilde{\check{X}}'_{2}\check{T}'_{2}}_{\alpha_{1}\alpha_{2}}
	\sigma_{\alpha_{2}}
	,
\end{align}
\begin{align}
	&\rho^{\hat{X}'_{3}\check{X}'_{3}\tilde{\hat{X}}'_{3}\tilde{\check{X}}'_{3}}
	=
	\sum_{\hat{X}_{3},\check{X}_{3},\hat{T}_{3},\check{T}'_{3}}
	\sum_{\alpha_{2}}
	\nonumber\\
	&
	\sigma_{\alpha_{2}}
	\Gamma^{\hat{X}_{3}\check{X}_{3}\hat{T}_{3}\hat{X}'_{3}\check{X}'_{3}\check{T}'_{3}}_{\alpha_{2}}
	\sigma_{\alpha_{2}}
	{\Gamma^{*}}^{\hat{X}_{3}\check{X}_{3}\hat{T}_{3}\tilde{\hat{X}}'_{3}\tilde{\check{X}}'_{3}\check{T}'_{3}}_{\alpha_{2}}
	.
\end{align}
Since the current Grassmann tensor network is translationally invariant on a lattice, these reduced density matrices and their EVDs followed by the SVD give us squeezers for the spatial direction. 
Denoting spacetime squeezers as $E$ for the forward subscript $\hat{X}_{f}\check{X}_{f}$, and $F$ for the backward subscript $\hat{X}'_{f}\check{X}'_{f}$, we have
\begin{align}
\label{eq:squz_e}
	E^{\hat{X}_{f}\check{X}_{f}X_{f}}
	=
	\sum_{x_{f}}
	(R^{-1})^{\hat{X}_{f}\check{X}_{f}x_{f}}
	U^{x_{f}X_{f}}\sqrt{\sigma_{X_{f}}}
	,
\end{align}
\begin{align}
\label{eq:squz_f}
	F^{\hat{X}'_{f}\check{X}'_{f}X'_{f}}
	=
	\sum_{x'_{f}}
	\sqrt{\sigma_{X'_{f}}}
	(V^{*})^{x'_{f}X'_{f}}
	(L^{-1})^{\hat{X}'_{f}\check{X}'_{f}x'_{f}}
	,
\end{align}
with $f=1,2,3$.
Note that $R$ and $L$ can be obtained from the EVDs of reduced density matrices and $U$, $V^{*}$. $\sigma$'s in Eqs.~\eqref{eq:squz_e} and \eqref{eq:squz_f} are given by the SVD of $RL$.
Unlike the flavor squeezers, we always take the square root of the singular values and they are assigned for each pair of squeezers equally. 
We refer to them as spacetime squeezers, whose derivation is similar to that of flavor squeezers, explained previously.
Note that when we derive spacetime squeezers, we need to carry out EVDs of reduced density matrices under their block diagonalized forms, which are helpful to update binary functions $f_{X}$ in Eq.~\eqref{eq:binary_func}. 
See Ref.~\cite{Akiyama:2021nhe} for detail.
The spacetime coarse-graining of $\Gamma$'s are accomplished by
\begin{align}
\label{eq:cg_x1}
	\Gamma^{X_{1}\hat{T}_{1}X'_{1}\check{T}'_{1}}_{\alpha_{1}}
	=
	\sum_{\hat{X}_{1},\check{X}_{1},\hat{X}'_{1},\check{X}'_{1}}
	\Gamma^{\hat{X}_{1}\check{X}_{1}\hat{T}_{1}\hat{X}'_{1}\check{X}'_{1}\check{T}'_{1}}_{\alpha_{1}}
	E^{\hat{X}_{1}\check{X}_{1}X_{1}}
	F^{\hat{X}'_{1}\check{X}'_{1}X'_{1}}
	,
\end{align}
\begin{align}
\label{eq:cg_x2}
	\Gamma^{X_{2}\hat{T}_{2}X'_{2}\check{T}'_{2}}_{\alpha_{1}\alpha_{2}}
	=
	\sum_{\hat{X}_{2},\check{X}_{2},\hat{X}'_{2},\check{X}'_{2}}
	\Gamma^{\hat{X}_{2}\check{X}_{2}\hat{T}_{2}\hat{X}'_{2}\check{X}'_{2}\check{T}'_{2}}_{\alpha_{1}\alpha_{2}}
	E^{\hat{X}_{2}\check{X}_{2}X_{2}}
	F^{\hat{X}'_{2}\check{X}'_{2}X'_{2}}
	,
\end{align}
\begin{align}
\label{eq:cg_x3}
	\Gamma^{X_{3}\hat{T}_{3}X'_{3}\check{T}'_{3}}_{\alpha_{2}}
	=
	\sum_{\hat{X}_{3},\check{X}_{3},\hat{X}'_{3},\check{X}'_{3}}
	\Gamma^{\hat{X}_{3}\check{X}_{3}\hat{T}_{3}\hat{X}'_{3}\check{X}'_{3}\check{T}'_{3}}_{\alpha_{2}}
	E^{\hat{X}_{3}\check{X}_{3}X_{3}}
	F^{\hat{X}'_{3}\check{X}'_{3}X'_{3}}
	,
\end{align}
where the size of $X_{f}$ with $f=1,2,3$ is fixed up to $D$, which is referred to as the spacetime bond dimension. 
Although the spacetime coarse-graining is apparently carried out for each $\Gamma$, each flavor, separately, this provides us with a possible coarse-graining of $(TT)$ in Eq.~\eqref{eq:combined_t}.
This is a direct benefit of constructing canonical forms for $(TT)$ employing flavor squeezers.
Redefining $T_{f}=\hat{T}_{f}$, $T'_{f}=\check{T}_{f}$ for $f=1,2,3$, we obtain the coarse-grained coefficient tensor, whose form is completely same with the right-hand side of Eq.~\eqref{eq:init_mps} (see Fig.~\ref{fig:cg_step_3}).
Thus, the procedure explained above can be easily repeated under the fixed flavor bond dimension $\chi$ and the spacetime bond dimension $D$.
At each coarse-graining step, the number of Grassmann tensors in Eq.~\eqref{eq:path_integral} is reduced to half.
Repeating the coarse-graining procedure alternately along the temporal and spatial directions, as the HOTRG does, one can evaluate the path integral of Eq.~\eqref{eq:path_integral} on the thermodynamic lattice. 
Although our explanation has assumed $N_{f}=3$, the procedure is easily applicable for the case with $N_{f}=2$.

Finally, we summarize the computational cost of the current coarse-graining procedure, which consists of three tasks, (i) to derive the flavor squeezers, (ii) to derive the spacetime squeezers, and (iii) the tensor contraction with these squeezers. 
Suppose we apply the method to the $d$-dimensional $N_{f}$-flavor lattice theory.
The computational cost of (i), dominated by tensor contraction to obtain reduced density matrices, scales with $\lo(\chi^{4}D^{{\mathrm{max}}(2d+1,2N_{f}-2)})$ with the $\lo(\chi^{2}D^{2(N_{f}-1)})$ memory cost.
The computational cost of (ii), also dominated by tensor contraction to obtain reduced density matrices, scales with $\lo(\chi^{4}D^{2d+2})$ with the $\lo(\chi^{4}D^{4})$ memory cost. 
The computational cost of (iii) scales with $\lo(\chi^{4}D^{4d-1})$ with the $\lo(\chi^{2}D^{2d})$ memory cost. 
The task of (iii) is similar to the coarse-graining in the $(d+1)$-dimensional HOTRG. 
Therefore, the computational cost of (iii) should scale with $\lo(\chi^{4}D^{4d-1})$.

\bibliography{bib/formulation,bib/algorithm,bib/discrete,bib/grassmann,bib/continuous,bib/gauge,bib/review,bib/for_this_paper}

\end{document}